\newif\ifanonymous
\anonymousfalse

\documentclass[a4paper,UKenglish,cleveref, autoref,thm-restate]{lipics-v2021}

\usepackage{graphicx} 
\usepackage{xspace}

\usepackage{siunitx}
\usepackage{float} 
\usepackage{amsmath,amssymb,amsfonts}
\usepackage{booktabs,caption,subcaption}
\usepackage{amsmath}
\usepackage{algorithm}
\newcommand{\norm}[1]{\left\lVert#1\right\rVert}

\usepackage{algorithmic}
\usepackage{graphics,graphicx,xcolor}
\usepackage[utf8]{inputenc}
\usepackage{diagbox}

\definecolor{honestblue}{RGB}{42,102,204}
\definecolor{adversaryred}{RGB}{204,32,32}
\definecolor{altblue}{RGB}{30,140,220}
\definecolor{bribeegreen}{RGB}{50,160,60}
\definecolor{evilorange}{RGB}{240,80,30}
\definecolor{evilgreen}{RGB}{51,184,64}

\newcommand{\cf}{\emph{cf.}\xspace}

\newcommand{\eg}{\textit{e.g.,} }
\newcommand{\ie}{\textit{i.e.,} }

\usepackage{listings}
\usepackage{listings-solidity} 

\usepackage{pdflscape}

\usepackage{xcolor}
\definecolor{verylightgray}{rgb}{.97,.97,.97}

\usepackage{listings, xcolor}

\definecolor{verylightgray}{rgb}{.97,.97,.97}

\lstdefinelanguage{Solidity}{
	keywords=[1]{anonymous, assembly, assert, balance, break, call, callcode, case, catch, class, constant, continue, constructor, contract, debugger, default, delegatecall, delete, do, else, emit, event, experimental, export, external, false, finally, for, function, gas, if, implements, import, in, indexed, instanceof, interface, internal, is, length, library, log0, log1, log2, log3, log4, memory, modifier, new, payable, pragma, private, protected, public, pure, push, require, return, returns, revert, selfdestruct, send, solidity, storage, struct, suicide, super, switch, then, this, throw, transfer, true, try, typeof, using, value, view, while, with, addmod, ecrecover, keccak256, mulmod, ripemd160, sha256, sha3}, 
	keywordstyle=[1]\color{blue}\bfseries,
	keywords=[2]{address, bool, byte, bytes, bytes1, bytes2, bytes3, bytes4, bytes5, bytes6, bytes7, bytes8, bytes9, bytes10, bytes11, bytes12, bytes13, bytes14, bytes15, bytes16, bytes17, bytes18, bytes19, bytes20, bytes21, bytes22, bytes23, bytes24, bytes25, bytes26, bytes27, bytes28, bytes29, bytes30, bytes31, bytes32, enum, int, int8, int16, int24, int32, int40, int48, int56, int64, int72, int80, int88, int96, int104, int112, int120, int128, int136, int144, int152, int160, int168, int176, int184, int192, int200, int208, int216, int224, int232, int240, int248, int256, mapping, string, uint, uint8, uint16, uint24, uint32, uint40, uint48, uint56, uint64, uint72, uint80, uint88, uint96, uint104, uint112, uint120, uint128, uint136, uint144, uint152, uint160, uint168, uint176, uint184, uint192, uint200, uint208, uint216, uint224, uint232, uint240, uint248, uint256, var, void, ether, finney, szabo, wei, days, hours, minutes, seconds, weeks, years},	
	keywordstyle=[2]\color{teal}\bfseries,
	keywords=[3]{block, blockhash, coinbase, difficulty, gaslimit, number, timestamp, msg, data, gas, sender, sig, value, now, tx, gasprice, origin},	
	keywordstyle=[3]\color{violet}\bfseries,
	identifierstyle=\color{black},
	sensitive=true,
	comment=[l]{//},
	morecomment=[s]{/*}{*/},
	commentstyle=\color{gray}\ttfamily,
	stringstyle=\color{red}\ttfamily,
	morestring=[b]',
	morestring=[b]"
}

\usepackage{url}
\usepackage{pst-node}
\usepackage{multido}

\usepackage{tikz}
\usetikzlibrary{positioning,chains,fit,shapes,calc}
\usepackage[most]{tcolorbox}
\usepackage{comment}
\usepackage[section]{placeins}

\usepackage[utf8]{inputenc}
\usepackage{pgfplots}
\DeclareUnicodeCharacter{2212}{−}
\usepgfplotslibrary{groupplots,dateplot,fillbetween}
\usetikzlibrary{patterns,shapes.arrows}
\pgfplotsset{compat=newest}

\definecolor{lowblue}{RGB}{0,114,178}
\definecolor{highorange}{RGB}{230,159,0}
\definecolor{allgray}{RGB}{153,153,153}

\usepackage{tikz}
\usetikzlibrary{arrows.meta,positioning,calc,fit}

\usepackage{pifont}

\tikzset{
  block/.style = {draw, thick, rounded corners=2pt, minimum width=1.7cm, minimum height=.8cm, align=center},
  honest/.style = {block, fill=honestblue!85, text=white},
  adversary/.style = {block, draw=adversaryred, dashed, very thick, fill=adversaryred!15, text=adversaryred},
  honestarw/.style = {altblue, -{Latex[length=3mm]}, very thick},
  advarw/.style = {adversaryred, dashed, -{Latex[length=3mm]}, thick},
  altarw/.style = {bribeegreen, dash dot, -{Latex[length=3mm]}, thick},
achain/.style = {black, dashed, -{Latex[length=3mm]}, thick},
chain/.style = {black, -{Latex[length=3mm]}, thick},
slotline/.style = {gray!55, line width=0.6pt},
}

\usepackage{makecell}

\usepackage{mathtools}

\newcommand{\row}{\mathsf{rowNum}}

\newcommand{\glabel}[1]{\{X^b\}}

\newcommand{\negl}{\mathsf{negl}}

\newcommand{\myparagraph}[1]{\vspace{.5em}\noindent\textbf{#1}}

\usepackage{xcolor}

\newif\ifrev
\revtrue   

\def\TikzAFTDataPath{Figures/}

\definecolor{corange}{RGB}{230,159,0}
\definecolor{csky}{RGB}{86,180,233}
\definecolor{cdeposit}{RGB}{0,158,115}
\definecolor{cwithdraw}{RGB}{213,94,0}
\definecolor{cteal}{RGB}{68,170,153}
\definecolor{cblue}{RGB}{0,114,178}
\definecolor{cgray}{RGB}{127,127,127}
\definecolor{cpurple}{RGB}{117,112,179}
\definecolor{darkgrey176}{RGB}{176,176,176}
\definecolor{green}{RGB}{0,128,0}
\definecolor{steelblue}{RGB}{70,130,180}
\definecolor{coral}{RGB}{255,127,80}
\definecolor{crimson}{RGB}{220,20,60}
\definecolor{navy}{RGB}{0,0,128}
\definecolor{forestgreen4416044}{RGB}{44,160,44}
\definecolor{darkorange}{RGB}{255,140,0}
\definecolor{skyblue}{RGB}{135,206,235}
\definecolor{mediumseagreen}{RGB}{60,179,113}
\definecolor{darkred}{RGB}{139,0,0}

\definecolor{coral}{RGB}{255,127,80}
\definecolor{darkgrey176}{RGB}{176,176,176}
\definecolor{lightgrey204}{RGB}{204,204,204}
\definecolor{purple}{RGB}{128,0,128}

\definecolor{lightgrey204}{RGB}{204,204,204}
\definecolor{crimson2143940}{RGB}{214,39,40}
\definecolor{darkorange25512714}{RGB}{255,127,14}
\definecolor{forestgreen4416044}{RGB}{44,160,44}
\definecolor{mediumpurple148103189}{RGB}{148,103,189}
\definecolor{steelblue31119180}{RGB}{31,119,180}

\newcommand{\figurewidth}{13cm}
\newcommand{\figureheight}{4cm}
\pgfplotsset{
  figurestyle/.style={
    width=\figurewidth,
    height=\figureheight,
    title style={font=\small},
    label style={font=\small},
    tick label style={font=\footnotesize},
    legend style={font=\small},
  },
  halfwidth/.style={
    width=6cm  
  },
}

\ifanonymous
\author{Invisible Authors}{Invisible Affiliations}{}{}{}
\else
\author{Kanan Huseynov}
{Department of Informatics, Eötvös Loránd University, Budapest, Hungary}
{kananhusayn@protonmail.com}
{}
{}

\author{Ali Shahzaib}
{Department of Informatics, Eötvös Loránd University, Budapest, Hungary}
{shahzaibaliofficial0@gmail.com}
{}
{}

\author{István András Seres}
{Department of Informatics, Eötvös Loránd University, Budapest, Hungary}
{seresistvanandras@gmail.com}
{}
{}

\author{János Tapolcai}
{Dept. of Telecommunications and 
AI, Budapest University of Technology and Economics, 
Hungary}
{tapolcai@tmit.bme.hu}
{}
{}
\fi

\Copyright{Jane Open Access and Joan R. Public} 

\ccsdesc[500]{Security and privacy~Pseudonymity, anonymity and untraceability}
\ccsdesc[300]{Security and privacy~Privacy-preserving protocols}
\ccsdesc[300]{Security and privacy~Economics of security and privacy}
\ccsdesc[100]{Applied computing~Digital cash}

\keywords{Ethereum, Blockchain, Anonymity, Privacy, Railgun} 

\category{} 

\relatedversion{} 

\nolinenumbers

\begin{document}

\title{\emph{A Tattered Cloak of Invisibility:}\enspace \\ Measuring Anonymity Loss in Railgun on Ethereum}

\ifanonymous
\authorrunning{Invisible Author et al.}
\else
\authorrunning{Huseynov et al.}
\fi

\Copyright{Copyright} 


\hypersetup{
    linkcolor=blue,
    filecolor=magenta,      
    urlcolor=blue,
    citecolor=blue,
    pdftitle={A Tattered Cloak of Invisibility: Measuring Anonymity Loss in Railgun on Ethereum} 
}

\EventEditors{John Q. Open and Joan R. Access}
\EventNoEds{2}
\EventLongTitle{42nd Conference on Very Important Topics (CVIT 2016)}
\EventShortTitle{CVIT 2016}
\EventAcronym{CVIT}
\EventYear{2016}
\EventDate{December 24--27, 2016}
\EventLocation{Little Whinging, United Kingdom}
\EventLogo{}
\SeriesVolume{42}
\ArticleNo{23}

\maketitle

\begin{abstract}
From a user’s perspective, perhaps the most significant difference between traditional banking services and widely used blockchain-based financial systems is that, in the latter, transactions and -- either directly or indirectly -- account balances and transaction histories are publicly observable. Therefore, a growing number of cryptographic solutions have been proposed to ``add a privacy layer'' to such systems. However, the privacy that users actually obtain does not depend solely on the security of the underlying cryptographic protocol: user behavior, transaction-amount patterns, and timing decisions can substantially reduce anonymity.

In this work, we study behavioral leakage in cryptocurrency mixers, focusing on Railgun on Ethereum. We aim to heuristically estimate the probability that a given deposit and withdrawal transaction belong to the same user. We consider five sources of leakage: characteristic timing patterns, address reuse, proximity in the transaction graph induced by prior public transactions, amount fingerprints that preserve distinctive digit patterns across transaction values, and knapsack-type matches in which groups of transaction amounts add up in revealing ways. Our results show that even cryptographically strong privacy systems may suffer substantial anonymity loss due to user behavior and transaction patterns. Our five heuristics are able to uniquely link $17.65\%$ of Railgun withdraw transactions to deposit transactions. We also applied a knapsack solver algorithm that was able to produce a $3.42$ bit median anonymity loss for withdraw transactions. This work contributes to a better understanding of the practical privacy limits of mixers and anonymity pools, and points toward safer usage practices and design principles.
\end{abstract}

\section{Introduction}

Cryptocurrencies have been remarkably successful as investment assets, yet have not fulfilled their original promise of becoming widely used money. This limitation is not merely technical. With the emergence of layer-2 systems, cryptocurrencies can in principle support low-cost, high-throughput, and even micropayment-scale transactions. A more fundamental barrier is related to privacy and user behavior. In traditional payments, the buyer and the merchant necessarily interact, and therefore they are not anonymous to each other. Precisely for this reason, it is essential that the payment process does not reveal the buyer's account history or balance to the merchant, nor the merchant's broader financial activity to the buyer. In contrast, widely used public blockchains expose transaction histories and, either directly or indirectly, account balances. As a result, a simple on-chain payment can reveal far more information than users would normally disclose in a traditional payment setting. For cryptocurrencies to serve as practical payment systems, they therefore need transaction mechanisms that preserve this basic form of financial privacy: neither party should learn the other party's transaction history or balance, while the merchant should still receive cryptographic assurance that the correct payment has been made.

At the same time, financial privacy creates an inherent regulatory tension. Privacy-preserving payment systems can also make it harder to detect tax evasion, money laundering, sanctions evasion, and other illicit activity. Tornado Cash, a privacy-enhancing overlay on Ethereum, is a prominent example~\cite{pertsev2019tornado}. In August 2022, the U.S. Department of the Treasury sanctioned a set of Tornado Cash smart-contract addresses, effectively prohibiting U.S. persons from interacting with them. Usage dropped sharply after the sanctions, and parts of the Ethereum ecosystem began to censor or avoid Tornado-Cash-related transactions~\cite{wahrstatter2024blockchain,wang2023blockchain}. The sanctions were later challenged in court, and in March 2025 the Treasury removed Tornado Cash from its sanctions list. Nevertheless, the episode illustrates the fragile position of privacy-enhancing infrastructure: it is simultaneously a tool for ordinary financial privacy and a source of serious compliance concerns.

Despite this growing need for privacy-preserving payment mechanisms, financial privacy is not a first-class default property in any of the top $10$ cryptocurrencies by market capitalization as of May 2026, according to CoinMarketCap. This has motivated a broad line of work aimed at adding a privacy layer to transparent blockchains, with stealth addresses, mixers and anonymity pools among the most prominent approaches~\cite{courtois2017stealth,meiklejohn2018mobius,pertsev2019tornado,seres2019mixeth}. At a high level, Ethereum-based versions of these systems are implemented as smart contracts that maintain a private pool of funds. Users deposit assets into the contract, and subsequent operations are represented as cryptographically hidden state transitions. Using cryptographic techniques, often based on zero-knowledge proofs, the system can verify that a transaction is valid without revealing which users' balances change, or by how much. The public blockchain observes that the pool state has been updated, but not the internal ownership changes that took place inside the pool. Eventually, users may withdraw funds from the private pool to a fresh public address. In a typical mixer, this means that a user first deposits funds into the pool and later withdraws them without publicly revealing which deposit funded the withdrawal. In principle, the identity of the withdrawer is hidden among all previous depositors, often referred as the corresponding anonymity set.

However, this cryptographic guarantee captures only one aspect of privacy. It prevents an observer from directly verifying the deposit--withdrawal link from the protocol transcript, but it does not eliminate all side information available on a public blockchain. Users may interact with the system in ways that leave behavioral traces: they may withdraw shortly after depositing, reuse related addresses, transfer distinctive amounts, or split and recombine funds in recognizable patterns. Such traces can shrink the effective anonymity set well below the nominal set of all prior deposits. This distinction between nominal and effective anonymity is crucial in practice. A mixer may provide a large cryptographic anonymity set, yet only a much smaller subset of deposits may be plausible candidates for a given withdrawal once timing, transaction amounts, and prior on-chain relationships are taken into account. In other words, privacy loss may arise not from a failure of the cryptographic protocol itself, but from the interaction between the protocol and user behavior.

In this work, we study this form of behavioral leakage in cryptocurrency mixers, focusing on Railgun on Ethereum as a case study. Railgun is particularly relevant because it extends the mixer model beyond simple fixed-denomination deposits/withdraws and supports more flexible private interactions inside the shielded pool. This flexibility is useful for users, but it also creates new opportunities for statistical and graph-based inference. Our goal is to estimate, using heuristic evidence, how likely a given deposit and withdrawal transaction belongs to the same user. As a solution to these privacy risks, various cryptographic solutions (cryptocurrency mixers and privacy pools) have been introduced,\eg Tornado Cash (TC). However, these solutions only provide limited features. Some limitations of these solutions include the following.
\begin{enumerate}
  \item \textbf{Limited Currencies:} Most mixers (most notably Tornado Cash) only support private transfer of native ETH and a handful of more ERC20 tokens. Railgun supports the deposit of every ERC20 tokens into the Railgun shielded pool.
  \item \textbf{Limited Functionalities:} TC only allowed the deposits and withdraws of cryptoassets. On the other hand, Railgun aims to provide a much richer sets of functionalities such as internal shielded  (thus, enabling private, confidential asset transfers), private decentralized finance applications (Uniswap, Compound, etc.). Thus, Railgun hopes to increase the anonymity sets and incentivize users to leave longer their funds in the shielded pool than as if there were no added financial functionalities.
\end{enumerate}

\subsection{Our contributions}
In this work, we provide the following contributions.

\begin{description}
    \item[Behavioral leakage perspective.]
    We study privacy in shielded-pool systems from a behavioral perspective. Rather than looking for cryptographic failures, we ask how much information remains available through public timing, amount, and transaction-graph patterns, and how this information can reduce the anonymity users obtain in practice.

    \item[Heuristics for linking deposits and withdrawals.]
    We introduce four families of heuristics for identifying plausible deposit--withdrawal links: address reuse, proximity in the public transaction graph, amount fingerprints, and knapsack-type amount matches. These heuristics are not intended to provide definitive deanonymization, but to estimate how much behavioral evidence points toward particular links.

    \item[Measurement study of Railgun on Ethereum.]
    We apply these heuristics to Railgun~\cite{buterin2024blockchain} on Ethereum and characterize the resulting anonymity loss. Our results show that even when cryptographic mechanisms hide direct links, observable user behavior can substantially narrow the set of plausible origins for withdrawals.

    \item[Empirical usage statistics.]
    We provide an empirical characterization of Railgun usage on Ethereum, including deposit and withdrawal behavior, timing patterns, transaction amounts, and the observable public-graph relationships around users entering and leaving the shielded pool.
    \item[Open-source code and datasets.] We release as an open-source repository the entire collected dataset, and the code we developed to analyze it. The repository is available at the following link:\ifanonymous ~\url{https://anonymous.4open.science/r/railgun_deanonymization-48A0}
    \else  ~\url{https://github.com/c0rt3x1337x/railgun_deanonymization}.
    \fi 
\end{description}

\subsection{Ethical considerations}\label{sec:ethics}
We study behavioral leakage in a deployed, publicly accessible privacy system. All data used in our analysis was collected exclusively from the public Ethereum blockchain, which is permissionlessly readable by any party without authentication or special access. We did not interact with the Railgun protocol in any active capacity, did not attempt to inject transactions, and did not access any off-chain data sources such as IP addresses, user accounts, or private communications. Our adversary model is therefore strictly passive and confined to publicly available information, consistent with the threat model described in~\Cref{sec:threat_model}.
 
Our analysis operates at the level of Ethereum addresses, not individuals. We do not resolve pseudonymous addresses to real-world identities and do not publish any mapping between addresses and persons. The introduced heuristics produce probabilistic links between deposit and withdraw \emph{addresses}; they do not identify the human beings who control those addresses. We deliberately refrain from publishing the list of flagged address pairs.
 
The anonymity losses we document arise from \emph{user behavior}, not from cryptographic vulnerabilities in Railgun's zero-knowledge proof system. There is therefore no technical vulnerability to disclose in the traditional sense. Nevertheless, we shared a preliminary version of our findings with the Railgun development team prior to submission, in line with responsible disclosure practice for privacy measurement studies. The heuristics we describe are passive, observable by any blockchain analyst, and do not require any special tooling or insider access. Publishing them advances the public understanding of practical privacy limits and enables users to make more informed decisions.

The rest of this paper is organized as follows. In~\Cref{sec:background}, we describe the necessary background on shielded pools on Ethereum, with a particular focus on Railgun. \Cref{sec:usage_statistics} studies behavioral and usage patterns in the Railgun ecosystem. We formulate our deanonymization heuristics in~\Cref{sec:heuristics} and evaluate them in~\Cref{sec:evaluation}. We review related work in~\Cref{sec:related_work}. Finally, we conclude with open research directions in~\Cref{sec:conclusion}.

\section{Background}\label{sec:background}

\subsection{Notations}\label{sec:notations}
To formalize the cryptographic and blockchain primitives discussed in this paper, we adopt the following notations: for a summary of our notations, see~\Cref{tab:notations}.

\begin{table}[tb]
    \centering
    \begin{tabular}{c|l}
       Notation & Description \\
       \midrule
       $\mathcal{D},\mathcal{W},\mathcal{I}$&Multiset of depositor, withdrawer, and internal transactions senders, respectively\\
       $\mathrm{supp}(\mathcal{D})$& The support of a (multi)set $\mathcal{D}$\\
       0x/0zk& Public/private Ethereum address types\\ 
       $d_i,w_j$ & The address that sent the $i$th deposit, and received the $j$th withdraw transaction \\
       $\mathcal{B}$ & A blocklist containing flagged, non-compliant deposit addresses \\
       $\mathsf{sender}(w_j)$ & The sender of the $j$th withdraw transaction \\  
       $t(d_i)$ & The timestamp when the $i$th deposit transaction had been sent\\
       $\mathsf{tx}(d_i,w_j)$& A boolean variable evaluating to $\mathsf{true}$ ($1$) iff there is a transaction sent from $d_i$ to $w_j$\\
       $\mathsf{amt}(d_i)$& Deposit amount of the $i$th (similarly, we allow $\mathsf{amt}$ to withdraw) deposit tx \\
       $d(x,y),\norm{x}$ & Hamming distance of two strings $x$ and $y$, Hamming-weight of $x$.
    \end{tabular}
    \caption{Notations used throughout the paper.}
    \label{tab:notations}
\end{table}

\subsection{Railgun: A Cloak of Invisibility}\label{sec:railgun}
Railgun is a privacy-enhancing layer on top of Ethereum implemented as a set of smart contracts. It is designed to allow users to deposit and withdraw funds from this private layer, often referred to as the Railgun shielded pool. Additionally, users can send confidential payments inside the shielded pool and can interact with  on-chain decentralized applications (dApp), \eg decentralized exchanges, lending pools, decentralized finance applications, etc., in a private manner. Railgun introduces a new address type on Ethereum called 0zk addresses.
\begin{description}
    \item[0x addresses] This is the native Ethereum address type that does not provide confidentiality, \ie any transaction to/from 0x addresses is visible on the blockchain with their transaction amounts exposed to the public. This allows any chain analyst to create a complete transaction history and accurate balance of every 0x address.
    \item[0zk addresses] A new private address type implemented by Railgun. Any transaction between two 0zk addresses is confidential, \ie the transaction amounts, and the sender/receiver 0zk addresses remain hidden to the public. Thus, in principle, no chain analyst can establish neither the balance nor the transaction history of any 0zk addresses.
\end{description}

Next, we review the actors and their possible actions on Railgun focusing on the privacy details and omit all unnecessary cryptographic details.
\begin{figure}
    \centering
    \includegraphics[width=0.9\linewidth]{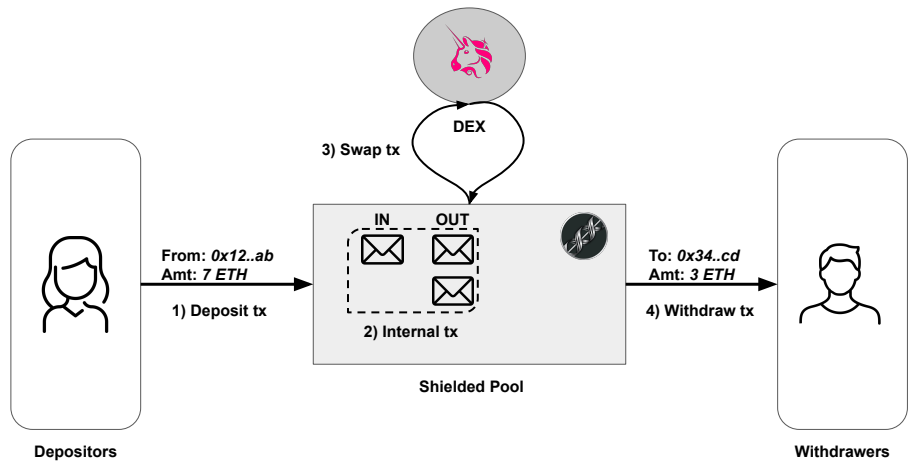}
    \caption{The Railgun privacy-enhancing overlay. Users can have four types of interactions with the Railgun smart contracts. Users can deposit (often shield) funds into the shielded pool. After funds are shielded, users can make internal transactions whose amounts are hidden, also known as confidential transactions. Users can use their shielded notes to unshield them and use them either in DeFi (\eg Uniswap) or withdraw (unshield) it back to the public Ethereum transaction graph.}
    \label{fig:railgun_system_model}
\end{figure}

\begin{description}
    \item[Depositors] Users can deposit $\mathsf{amt}(d_i)$ funds to any 0zk address within the shielded pool by issuing a deposit transaction $d_i$ at time $t(d_i)$. Often we refer both to the public 0x sender address and the deposit transaction as $d_i$. Users may issue many deposit transactions. 
    \item[Railgun smart contract] Every Railgun functionality (\ie deposit, withdraw, and internal transactions) requires users to issue transactions with the Railgun smart contract as the recipient address. It stores a Merkle tree of cryptographic notes and nullifiers. 
    \item[Withdrawers] An owner of a cryptographic note in the shielded pool may withdraw that note with value $\mathsf{amt}(w_j)$ back to the public 0x Ethereum address. Our primary focus in this work is to understand which withdrawer $w_j$ can be linked to a previous depositor $d_i$.
    \item[Users] Note that a Railgun user is not necessarily a depositor or withdrawer. One may receive and send funds inside the shielded pool without depositing or withdrawing them from the shielded pool,~\ie one can send and receive funds from a 0zk address they own.
    \item[Relayers] When a user wants to withdraw funds to a fresh address $w_j$, typically $w_j$ lacks funds to cover the transaction fees of the withdraw transaction. Thus, a third party may submit the withdraw transaction on behalf of the user (in exchange for a relayer fee) to avoid any unnecessary behavioural leakage. We identified $124$ relayers,~\cf~\Cref{fig:broadcaster_cluster}.
\end{description}

\begin{figure}[t!]
\begin{tikzpicture}
  \begin{axis}[
    figurestyle,
    date coordinates in=x, xmax=2026-04-26, xticklabel={\year}, xtick distance=1*365.25,
    xlabel={Year}, ylabel={Events/week},
    legend style={at={(0.02,0.98)}, anchor=north west},
    ymajorgrids, ymin=0,
  ]
    \addplot[cdeposit, thick] table
      [col sep=comma, x=week_start, y=shields]
      {\TikzAFTDataPath figure_ch4_02_weekly_boundary_counts.csv};
    \addlegendentry{Deposit}
    \addplot[cwithdraw, thick, dotted] table
      [col sep=comma, x=week_start, y=unshields]
      {\TikzAFTDataPath figure_ch4_02_weekly_boundary_counts.csv};
    \addlegendentry{Withdraw}
  \end{axis}
\end{tikzpicture}
\caption{Railgun weekly deposit and withdraw transaction counts on Ethereum L1.}
\label{fig:shield_unshield_counts}
\end{figure}
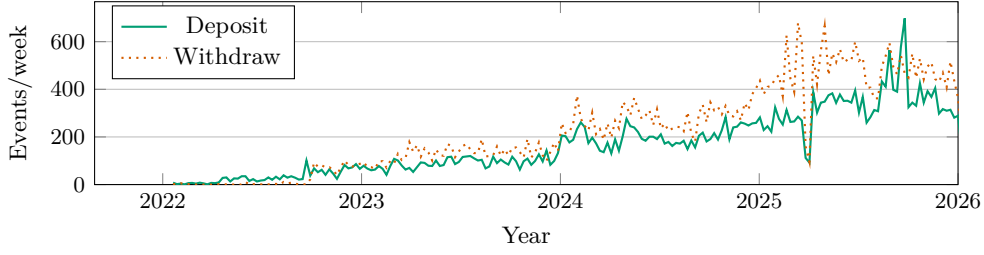

Railgun users may issue the following types of transaction.
\begin{description}
    \item[Deposit (shield) transaction] As of May 12, 2026, \num{122477} deposit transactions have been issued with a total volume of \$\num{5058127358} (Ethereum L1 and Arbitrum, Polygon, BSC). Importantly, Railgun applies a zero-knowledge compliance monitoring tool called Private Proofs of Innocence (PPOI). Every deposit transaction undergoes this analysis: if the depositor address $d_i$ is on a blocklist, the deposited funds cannot be used inside the shielded pool. Subsequently, Railgun only allows one to withdraw the blocklisted deposited amount back to the very same deposit address $d_i$. 
    \item[Internal transaction] Confidential transactions that occur inside the shielded pool between 0zk addresses are called internal transactions. A public observer only sees the number of input notes redeemed, and the number of newly created output notes,~\cf~\Cref{fig:number_of_input_notes}. 
    
    \item[Withdraw (unshield) transaction] As of May 12, 2026, \num{328520} withdraw transactions had been issued,~\cf~\Cref{fig:shield_unshield_counts}.
    \item[Swap transaction] Users may unshield some of their shielded notes and use them in various DeFi applications, \eg swapping assets on decentralized exchanges. 
    This is a moderately used feature in Railgun, \eg $\num{3133}$ swap transactions at the time of writing. For comprehensive usage statistics on Railgun swap transactions see~\Cref{fig:swap_tx_statistics}.
\end{description}

\subsection{Anonymity Guarantees and Their Measurement}\label{sec:privacy_guarantees}
In this work, we are primarily interested in breaking the following privacy guarantee provided by Railgun. Intuitively, every prior deposit transaction should act as an equally likely source for a given withdraw transaction. Next, we formalize this intuition. 
\begin{definition}(Deposit-withdraw transaction unlinkability)\label{def:dep-with-unlinkability} This unlinkability privacy guarantee requires that for every PPT adversary $\mathcal{A}$, and for a withdraw transaction $w_j$ that received (some of) its money from deposit transaction $d_i$, $\mathcal{A}$ has a negligible advantage in finding the real source $d_i$ for withdraw $w_j$. More formally, 
\begin{equation}\label{def:adversary_advantage}
    \forall d_i\in D(w_j):\Big\vert\Pr[\mathcal{A}(D(w_j),w_j)=d_i]-\frac{1}{\vert D(w_j)\vert}\Big\vert\leq\negl(\lambda)\enspace,
\end{equation}
    where $D(w_j):=\{d_i\vert t(d_i)\leq t(w_j)\}$, \ie is the set of deposit transaction issued before $t(w_j)$.
\end{definition}
In a similar vein, one could define deposit-internal transaction unlinkability, deposit-swap transaction, or internal-withdraw transaction unlinkability, etc. We leave it to future work to define, assess and potentially break these privacy properties in Railgun. 

The adversary's goal is to determine which deposit $d_j$ can be linked to a withdraw $w_i$. To that end, the adversary outputs a discrete probability distribution over $D(w_i)$. Thus, we want to measure the uncertainty of a given adversary $\mathcal{A}$ about the sources of funds for a withdraw transaction $w_j$.
\begin{definition}(Entropic anonymity measure~\cite{diaz2002towards,serjantov2002towards})\label{def:anonymity_measure}
For a withdraw $w_j$ and adversary $\mathcal{A}$, let $X_{j,\mathcal{A}}$ denote the probability distribution over $D(w_j)$ \ie, for $d_i\in D(w_j)$, we have $\Pr[X_{j,\mathcal{A}}\in \{d_i\}]:=p_i$ . Let the anonymity measure $\mathbb{A}(w_j)$ of $w_j$ be the Shannon-entropy of $X_{j,\mathcal{A}}$. Formally, we have     
\begin{equation}
   \mathbb{A}(w_j):=H(X_{j,\mathcal{A}})= -\sum_{d_i\in D(w_j)}p_i\log_2(p_i)\enspace.
\end{equation}
    
\end{definition}

In the context of Railgun, when one wants to break deposit-withdraw unlinkability,~\cf~\Cref{def:dep-with-unlinkability}, the ideal anonymity measure that Railgun achieves would be that for each withdraw transaction $w_j$, the adversary deems each prior $d_i$ equally probable. We call this the optimistic anonymity guarantee achieved by $w_j$
\begin{definition}(Naïve or optimistic anonymity)\label{def:naive_anonymity} We define an idealistic upper bound on the anonymity measure of $w_j$ as $\mathbb{A}(w_j):=\log_2(\vert D(w_j) \vert)$.
\end{definition}
The main goal of this work is to define adversarial strategies that give a more realistic upper bound than the optimistic Shannon-entropic measure on the achieved anonymity guarantees of Railgun. We focus on measuring anonymity using the classical Shannon entropy-based anonymity measures defined above. We leave it to future work to apply and evaluate other anonymity measures~\cite{wagner2018technical} for assessing the anonymity provided by Railgun.

\subsection{Threat Model}\label{sec:threat_model}
We assume a passive adversary, \ie we only assume access to the public Ethereum blockchain data. This adversary may be \eg a blockchain analyst. In our analysis, we do not use metadata leakages that come from other layers of the decentralized technology stack, \eg network-level leakages~\cite{wang2025time}, or other public data sources (\eg social media). However, we note that, for example, relayers or other Ethereum full nodes could link the depositor/withdrawer IP addresses to their public 0x Ethereum addresses. We leave it to future work to evaluate the efficacy of active attacks on privacy, for example, one could launch an active amount fingerprinting attack inside the shielded pool~\cite{biryukov2019privacy}. In this attack, the adversary sends a maliciously crafted transaction in the hope of altering certain digits in the 0zk balance of an address. If at withdrawal time those digits remain unchanged, then the adversary could follow the flow of funds back to the public part of Ethereum (\ie 0x addresses).
\section{Railgun Usage Statistics}\label{sec:usage_statistics}
\subsection{Dataset}\label{sec:dataset}
\begin{figure}[t!]
  \centering

\begin{tikzpicture}
  \begin{axis}[
    width=0.7*\figurewidth, height=\figureheight,
    date coordinates in=x, xticklabel={\year},
    xtick distance=1*365.25, date ZERO=2019-01-01,
    ymin=-0.6, ymax=5.6,
    ytick={0,1,2,3,4,5},
    yticklabels={{Railgun Shield events}, {Railgun Unshield events (V2)}, {Railgun Internal transfers}, {Railgun RelayAdapt swaps}, {External ETH transfers}, {External ERC20 transfers}},
    y dir=reverse,
    xlabel={Year},
    enlarge x limits=0.02,
    xmajorgrids,
  ]
    \addplot[cdeposit, line width=8pt, mark=|, mark size=6pt, mark options={line width=1pt}] coordinates {(2022-05-13,0) (2026-04-29,0)};
    \addplot[cwithdraw, line width=8pt, mark=|, mark size=6pt, mark options={line width=1pt}] coordinates {(2022-05-15,1) (2026-04-29,1)};
    \addplot[cblue, line width=8pt, mark=|, mark size=6pt, mark options={line width=1pt}] coordinates {(2022-05-11,2) (2026-04-29,2)};
    \addplot[corange, line width=8pt, mark=|, mark size=6pt, mark options={line width=1pt}] coordinates {(2022-12-02,3) (2025-12-24,3)};
    \addplot[csky, line width=8pt, mark=|, mark size=6pt, mark options={line width=1pt}] coordinates {(2019-06-12,4) (2026-02-09,4)};
    \addplot[cteal, line width=8pt, mark=|, mark size=6pt, mark options={line width=1pt}] coordinates {(2019-06-03,5) (2026-02-09,5)};
    \draw[dashed, thick, cgray] (axis cs:2023-01-01,\pgfkeysvalueof{/pgfplots/ymin})
      -- (axis cs:2023-01-01,\pgfkeysvalueof{/pgfplots/ymax})
      node[pos=0.04, anchor=south west, font=\scriptsize, fill=white,
           inner sep=2pt] {H2 post-2023 cutoff};
  \end{axis}
\end{tikzpicture}
  \caption{The coverage of our collected datasets.}\label{fig:data_coverage}
\end{figure}
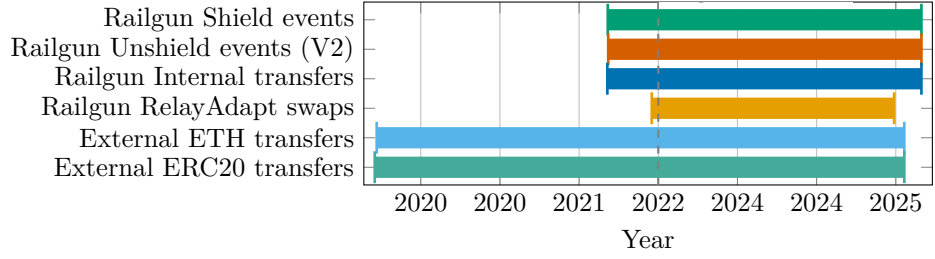

We collected every transaction of Railgun users in relation to deposit, withdraw, internal, and swap transactions,~\cf~\Cref{fig:data_coverage}. In addition, we also collected every transaction sent from all identified Railgun users. We mostly focused on their ETH and ERC20 token transfers.
\subsection{Limitations}\label{sec:limitations}
We solely focus on the Ethereum Layer-1 blockchain, though Railgun has been deployed on other Layer-2 networks as well, such as Arbitrum, Polygon, and Binance Smart Chain. Moreover, this work focuses solely on the native currency ether, even though Railgun supports the shielding and unshielding of ERC-20 tokens (\eg USDC, USDT, DAI, etc.) as well. 
\subsection{Railgun Usage}\label{sec:railgun_usage}
\begin{figure}
\centering
\subfloat[\label{fig:cumulative_shield_volumes}
Cumulative WETH shielded and unshielded over time. 
The retained balance, i.e., the difference between the two curves, is shown on the right.]{  

\begin{tikzpicture}
  \begin{axis}[
    figurestyle,halfwidth,
    date coordinates in=x, xticklabel={\year}, xtick distance=1*365.25,
    xlabel={Year}, ylabel={Cumulative WETH},
    legend style={at={(0.02,0.98)}, anchor=north west},
    ymajorgrids, ymin=0,
  ]
    \addplot[name path=shield, cdeposit, thick] table
      [col sep=comma, x=date, y=cum_shield]
      {\TikzAFTDataPath figure_ch4_03_cumulative_pool_flow.csv};
    \addlegendentry{Shield}

    \addplot[name path=unshield, cwithdraw, thick, dashed] table
      [col sep=comma, x=date, y=cum_unshield]
      {\TikzAFTDataPath figure_ch4_03_cumulative_pool_flow.csv};
    \addlegendentry{Unshield}

    \addplot[csky, fill opacity=0.20] fill between[of=shield and unshield];
  \end{axis}
\end{tikzpicture}\qquad\quad
}\enspace
\subfloat[\label{fg:retained}
Retained WETH balance (left axis) and FIFO-style residence time estimates (right axis) derived from cumulative shield and unshield flows.]{
\begin{tikzpicture}
  \begin{axis}[
    figurestyle,halfwidth,
    date coordinates in=x,
    xticklabel={\year},
    xtick distance=1*365.25,
    xlabel={Year},
    ylabel={Retained WETH},
    axis y line*=left,
    axis x line*=bottom,
    ymajorgrids,
    ymin=0, ymax=120000, xmin=2023-06-01,
    legend style={at={(0.02,0.98)}, anchor=north west,font=\footnotesize},
  ]
    \addplot[csky, thick] table
      [col sep=comma, x=date, y=retained]
      {\TikzAFTDataPath figure_ch4_03_little_law_timeseries.csv};
    \addlegendentry{$\mathrm{shield}(t)-\mathrm{unshield}(t)$}
  \end{axis}
  \begin{axis}[
    figurestyle,halfwidth,
    date coordinates in=x,
    xticklabel={\year},
    xtick distance=1*365.25,
    xlabel={Year},
    ylabel={Residence time [days]},
    axis y line*=right,
    axis x line=none,
    ymin=0, ymax=365, xmin=2023-06-01,
    legend style={at={(0.9,0.28)}, anchor=north east,font=\footnotesize},
  ]
    \addplot[cwithdraw, thick, dotted,
    ] table
      [col sep=comma, x=date, y=fifo_horizontal_lag_days]
      {\TikzAFTDataPath figure_ch4_03_little_law_timeseries.csv};
    \addlegendentry{FIFO-style estimate}
  \end{axis}
\end{tikzpicture}
}
  \caption{Cumulative shield and unshield volumes and implied residence time estimates in Railgun.}
\end{figure}
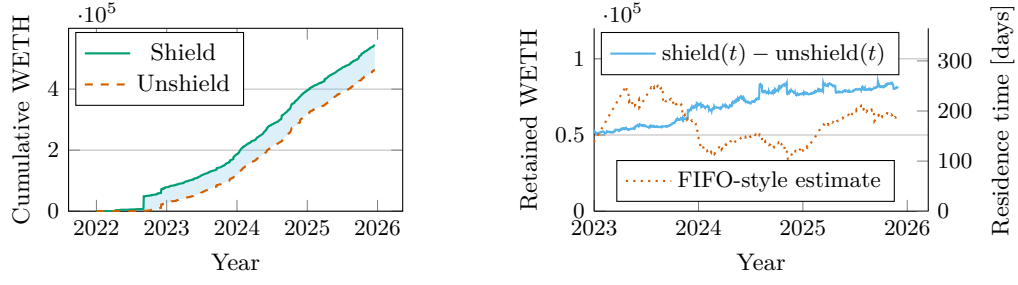

First, we study Railgun users' behavior at large. Most importantly, we measure how much money is left and used in the shielded pool. We observe in~\Cref{fig:cumulative_shield_volumes}, that as of 29th April, 2026, only $\num{81734}$ ETH has been left in the shielded pool, that is $15.00\%$ of the total cumulative deposits in Railgun's history. This suggests that the vast majority of Railgun users still treat Railgun's shielded pool as a simple Tornado Cash-like cryptocurrency mixer that has only two functionalities: deposits and withdraws.


\begin{figure}

\newcommand{\pdfscale}{495}

\begin{tikzpicture}
\begin{axis}[
  figurestyle,
  tick align=outside,
  tick pos=left,
  title={},
  xlabel={Time difference [days]},
  ylabel={Probability},
  x grid style={darkgrey176},
  y grid style={darkgrey176},
  xmajorgrids,
  ymajorgrids,
  xmin=0.001,ymax=1,
  scaled y ticks=false,
  xmode=log,
  ymode=log,
xtick={0.0416667,0.5,1,7,30,183,365},
  xticklabels={1 h,12 h,1 d,1 week,1 month,183 d,1 year},
  xticklabel style={
    font=\scriptsize,
    rotate=35,
    anchor=east
  },
]
\addplot [semithick, navy]
table [
  x=days,
  y expr=\thisrow{pdf}/\pdfscale,
  col sep=comma
] {Figures/cdf_deltas_pdf_data_sampled.csv};


\end{axis}
\end{tikzpicture}
  \caption{The probability density function of the time differences ($t(w_j)-t(d_i)$) between deposits $d_i$ and withdrawals $w_j$ linked by Heuristic 1 (deposit-withdraw address reuse),~\cf~\Cref{sec:heuristic1}.}
  \label{fig:cdf_deltas}
\end{figure}
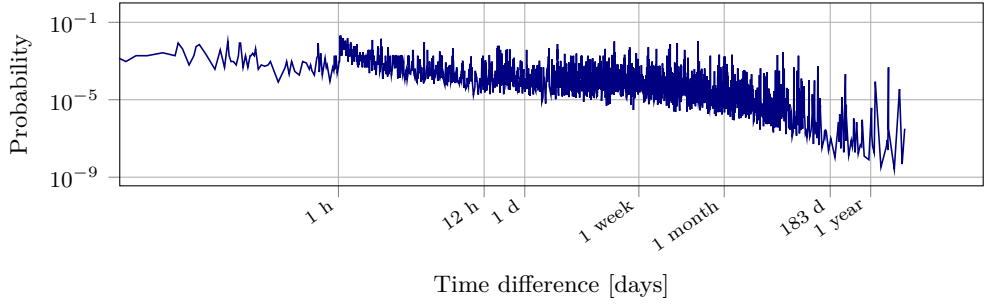

We can derive a rough estimate for the average residence time of funds in the Railgun pool using a Little's law style approximation~\cite{little1961proof}. Let $\overline{R}$ denote the average retained balance in the pool and let $\lambda_{\mathrm{out}}$ denote the average unshielding rate (ether/day). Under the assumption that the system is approximately in steady state over the observation window, the average residence time can be estimated per Little's theorem as
\begin{equation}
    T \approx \frac{\overline{R}}{\lambda_{\mathrm{out}}}
    = \frac{58\,633\text{\ WETH}}{320\text{\ WETH/day}}
    = 183 \text{ days}.
\end{equation}

Using the observed cumulative unshield volume and the length of the observation period, we estimate an average outflow of roughly $320$ WETH per day. Combined with an average retained balance of approximately $58\,633$ WETH (see ~\Cref{fg:retained}), this suggests that the typical residence time of funds inside the Railgun pool is on the order of several months.

~\Cref{fg:retained} also shows a FIFO-style estimate based on the horizontal lag between cumulative shield and cumulative unshield curves, i.e., the delay required for unshielded volume to ``catch up'' with prior shielded funds. This independent estimate yields a similar timescale, supporting the same several-month residence-time conclusion.


\section{Heuristics: Tattering the Cloak of Invisibility}\label{sec:heuristics}
Next, we formulate our heuristics (\cf~\Cref{fig:heuristic_illustration} for a figurative illustration) that we empirically evaluate in~\Cref{sec:evaluation} to link withdraw to deposit transactions. A trivial observation that we do not formulate as a separate heuristic is that a withdraw transaction $w_i$ cannot be linked to deposit transactions $d_j$ that occurred after $w_i$, \ie $\nexists (w_i,d_j): t(d_j)\geq t(w_i)$.
\subsection{Heuristic 1: deposit-withdraw address-reuse}\label{sec:heuristic1}
Ethereum's account-based ledger model encourages address reuse for better usability. Thus, in accordance with prior works~\cite{beres2021blockchain,wang2023zero,wu2022tutela}, we formulate our first heuristic.
\begin{definition}(Heuristic 1 (\textbf{H1})) If there exists $(i,j)\in\mathbb{N}^2$ such that $d_i=w_j\land t(d_i)\leq t(w_j)$, then we consider the deposit $d_i$ and withdraw $w_j$ transactions linked. 
\end{definition}
This is our only heuristic that, by definition, cannot produce false positives.

\subsection{Heuristic 2: direct transactional linkage}\label{sec:heuristic2}
Many users falsely believe that withdraw addresses are ``clean'' addresses as their direct source of funds is the ``clean'' shielded pool. Thus, whenever they run out of liquidity at a withdraw address $w_i$, they top its balance up by sending money from their other Ethereum accounts, often from a deposit address $d_j$,~\cf~\Cref{fig:monthly_deposit_withdraw_heuristic2_txs}. If a transaction occurs in either direction ($d_i\rightarrow w_j\lor w_j\rightarrow d_i$), we consider those deposit and withdraw transactions issued by the same user.
\begin{definition}(Heuristic 2 (\textbf{H2}))
    If $\exists(i,j)\in\mathbb{N}^2:\mathsf{tx}(d_i,w_j)=1\lor\mathsf{tx}(w_j,d_i)=1$, then we consider the deposit addresses $d_i$ and withdraw $w_j$ linked.
\end{definition}
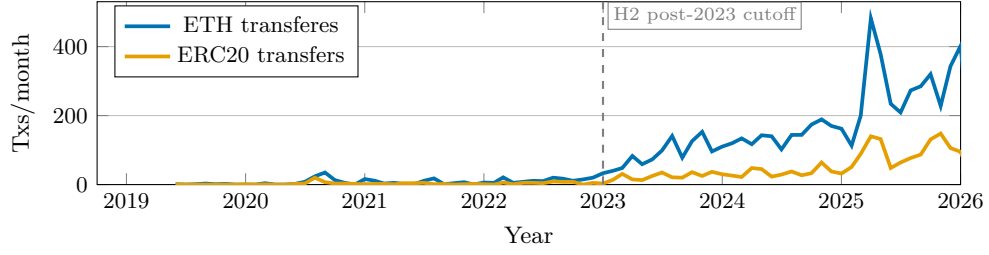
\begin{figure}[t!]
  \centering
\begin{tikzpicture}
  \begin{axis}[
    figurestyle,
    date coordinates in=x,  xmax=2026-01-1, xticklabel={\year}, xtick distance=1*365.25,
    xlabel={Year}, ylabel={Txs/month},
    ymajorgrids, ymin=0,
    legend style={at={(0.02,0.98)}, anchor=north west, font=\footnotesize},
  ]
    \addplot[cblue, line width=1.4pt] table
      [col sep=comma, x=month, y=eth_edges]
      {\TikzAFTDataPath figure_ch4_09_h2_monthly_edge_timeline.csv};
    \addlegendentry{ETH transferes}
    \addplot[corange, line width=1.4pt] table
      [col sep=comma, x=month, y=erc20_edges]
      {\TikzAFTDataPath figure_ch4_09_h2_monthly_edge_timeline.csv};
    \addlegendentry{ERC20 transfers}
    \draw[dashed, cgray, thick] (axis cs:2023-01-01,0)
      -- (axis cs:2023-01-01,\pgfkeysvalueof{/pgfplots/ymax});
    \node[anchor=north west, font=\scriptsize, fill=white,
          inner sep=2pt, draw=cgray, text=cgray]
      at (axis cs:2023-01-15,\pgfkeysvalueof{/pgfplots/ymax})
      {H2 post-2023 cutoff};
  \end{axis}
\end{tikzpicture}
  \caption{Number of monthly transactions between Railgun depositor and withdrawer addresses. The prevalence of these transactions motivates our formulation of Heuristic 2,~\cf~\Cref{sec:heuristic2}.}
  \label{fig:monthly_deposit_withdraw_heuristic2_txs}
\end{figure}
We acknowledge that this heuristic may introduce false positive links. Intuitively, Heuristic 2 exploits the connectedness of the deposit and withdraw addresses in the transaction graph. One could generalize Heuristic 2 such that it also considers deposit and withdraw addresses being linked if they are $k$-hop away from each other in the transaction graph (\eg for $k=1$ think of a swap transaction on a decentralized exchange with different in and out addresses). We did not pursue this direction, as it would have introduced more false positives. Alternatively, one could generalize this heuristic where the link between $d_i$ and $w_j$ is probabilistic and weighted by the number of transactions between the two addresses.

\subsection{Heuristic 3: withdraw transaction gas payer address reuse}\label{sec:heuristic_3_withdraw_gas_payer}
Currently in Ethereum, each transaction's sender must pay the transaction fee in ether, Ethereum's native cryptocurrency~\cite{seres2020blockchain}. However, when a user wants to withdraw funds from Railgun to a fresh withdraw address $w_j$ that had not been funded, they cannot send the withdraw transaction from $w_j$. Therefore, the withdraw transaction sender must be another \emph{unlinked} address. Railgun provides two solutions to this privacy-usability dilemma.
\begin{description}
    \item[Relayed withdraw transactions] Withdrawers may use a third-party, a so-called relayer, to send their withdraw transactions in exchange for a relayer fee. From a privacy point of view, this is the safest solution, since the relayer's address is typically independent from the user's withdraw address $w_j$.
    \item[Self-broadcast withdraw transactions] Users may opt out from using relayers and broadcast their withdraw transaction from another address they own. We denote this address as $\mathsf{sender}(w_j)$. If the address $\mathsf{sender}(w_j)$ can be linked in some way to a depositor address $d_i$, then we can also link $w_j$ to $d_i$.
\end{description}
Thus, we formulate our next heuristic as follows.
\begin{definition}(Heuristic 3 (\textbf{H3})) If the sender address of a withdraw transaction $\mathsf{sender}(w_i)\in\mathcal{D}$ and $\mathsf{sender}(w_i)$ itself is not a relayer, then we link $w_i$ and $\mathsf{sender}(w_i)$.  
\end{definition}
We detail in~\Cref{sec:heuristic3_evaluation}, how we differentiate between ``regular'' (\ie a user-relayed) and a relayed withdraw transactions. In other words, we need reliable and robust ways to identify relayers to reduce false positives produced by this heuristic.

\usetikzlibrary{positioning,calc,arrows.meta,backgrounds,fit}
\begin{figure}[t]
\centering
\begin{tikzpicture}[
    font=\small, >=Stealth,
    addr/.style={draw, rounded corners=2pt,
                 minimum width=1.05cm, minimum height=0.46cm,
                 inner sep=2pt, align=center,
                 font=\footnotesize\ttfamily},
    dep/.style={addr, fill=honestblue!18, draw=honestblue!75!black},
    wit/.style={addr, fill=adversaryred!12, draw=adversaryred!65!black},
    pool/.style={draw=black!40, fill=black!5, rounded corners=3pt,
                 minimum width=1.3cm, minimum height=0.52cm,
                 align=center, font=\scriptsize\itshape},
    topool/.style={->, thick, draw=black!40,
                   shorten >=1pt, shorten <=1pt},
    ev/.style={->, semithick, dashed,
               shorten >=2pt, shorten <=2pt},
    htitle/.style={font=\bfseries\footnotesize, align=center},
    panel/.style={rounded corners=5pt, draw=#1!30,
                  fill=#1!5, inner sep=8pt},
]

\def\dx{0}      
\def\px{1.65}   
\def\wx{3.3}    

\def\Aa{0}
\def\Ab{4.7}
\def\Ac{9.4}

\def\Ba{1.7}
\def\Bb{7.3}

\def\drop{-2.6}


\begin{scope}[xshift=\Aa cm]
  \node[dep]  (h1d) at (\dx,0) {$d_i$};
  \node[pool] (h1p) at (\px,0) {pool};
  \node[wit]  (h1w) at (\wx,0) {$w_j$};
  \draw[topool](h1d)--(h1p);
  \draw[topool](h1p)--(h1w);
  \draw[ev, draw=honestblue!75!black, bend right=45]
      (h1d.south) to
      node[below, font=\scriptsize, text=honestblue!70!black] {$d_i = w_j$}
      (h1w.south);
  \begin{scope}[on background layer]
    \node[panel=honestblue, fit=(h1d)(h1p)(h1w)] (h1box) {};
  \end{scope}
  \node[htitle, text=honestblue!85!black, above=4pt of h1box]
      {H1: Address Reuse};
\end{scope}

\begin{scope}[xshift=\Ab cm]
  \node[dep]  (h2d) at (\dx,0) {$d_i$};
  \node[pool] (h2p) at (\px,0) {pool};
  \node[wit]  (h2w) at (\wx,0) {$w_j$};
  \draw[topool](h2d)--(h2p);
  \draw[topool](h2p)--(h2w);
  \draw[ev, draw=adversaryred!75!black, bend right=45,<->]
      (h2d.south) to
      node[below, font=\scriptsize, text=adversaryred!70!black] {\textbf{on-chain tx}}
      (h2w.south);
  \begin{scope}[on background layer]
    \node[panel=adversaryred, fit=(h2d)(h2p)(h2w)] (h2box) {};
  \end{scope}
  \node[htitle, text=adversaryred!85!black, above=4pt of h2box]
      {H2: Direct Tx Link};
\end{scope}

\begin{scope}[xshift=\Ac cm]
  \node[dep]  (h3d) at (\dx,0) {$d_i$};
  \node[pool] (h3p) at (\px,0) {pool};
  \node[wit]  (h3w) at (\wx,0) {$w_j$};
  \draw[topool](h3d)--(h3p);
  \draw[topool](h3p)--(h3w);
  \node[dep, minimum width=0.9cm] (h3s) at (\px+1.3,0.9) {$d_i$};
  \node[font=\scriptsize, text=bribeegreen!80!black,
        left=3pt of h3s] {\textbf{gas payer}};
  \draw[ev, draw=bribeegreen!80!black](h3s.south)--($(h3p.north)+(0.8,-0.25)$);
  \begin{scope}[on background layer]
    \node[panel=bribeegreen, fit=(h3d)(h3p)(h3w)(h3s)] (h3box) {};
  \end{scope}
  \node[htitle, text=bribeegreen!85!black, above=4pt of h3box]
      {H3: Gas Payer Reuse};
\end{scope}

\draw[black!15] (-0.5,\drop+1.25) -- (12.9,\drop+1.25);


\begin{scope}[xshift=\Ba cm, yshift=\drop-3.6cm]
  \node[dep]  (h4d1) at (\dx, 0.32)  {$d_{i_1}$};
  \node[dep]  (h4d2) at (\dx,-0.32)  {$d_{i_2}$};
  \node[pool] (h4p)  at (\px, 0)     {pool};
  \node[wit]  (h4w)  at (\wx, 0)     {$w_j$};
  \draw[topool](h4d1.east)--(h4p.west);
  \draw[topool](h4d2.east)--(h4p.west);
  \draw[topool](h4p)--(h4w);
  \node[draw=evilorange!50, fill=evilorange!8, rounded corners=2pt,
        font=\scriptsize, align=center, inner sep=4pt]
      (h4ann) at (\px,1.05)
      {$\mathsf{amt}(d_{i_1})+\mathsf{amt}(d_{i_2})\approx\mathsf{amt}(w_j)$};
  \draw[ev, draw=evilorange!75!black](h4ann.south)--(h4p.north);
  \begin{scope}[on background layer]
    \node[panel=evilorange,
          fit=(h4d1)(h4d2)(h4p)(h4w)(h4ann)] (h4box) {};
  \end{scope}
  \node[htitle, text=evilorange!85!black, above=4pt of h4box]
      {H4: Knapsack Matching};
\end{scope}

\begin{scope}[xshift=\Bb cm, yshift=\drop-3.3cm]
  \node[dep]  (h5d) at (\dx,0) {$d_i$};
  \node[pool] (h5p) at (\px,0) {pool};
  \node[wit]  (h5w) at (\wx,0) {$w_j$};
  \draw[topool](h5d)--(h5p);
  \draw[topool](h5p)--(h5w);
  \node[font=\scriptsize\ttfamily, align=center]
      (h5fd) at (\px, 0.85)
      {$\mathsf{amt}(d_i)$:\ 1.\underline{\bfseries 345}67{\tiny\ldots}};
  \node[font=\scriptsize\ttfamily, align=center]
      (h5fw) at (\px,-0.85)
      {$\mathsf{amt}(w_j)$:\ 0.\underline{\bfseries 345}11{\tiny\ldots}};
  \node[font=\scriptsize, text=altblue!80!black]
      (matchlbl) at (\px+2.85,0) {\bfseries match};
  \draw[altblue!75!black, semithick]
      ($(h5fd.east)+(0.75,0)$) -- ++(0.22,0)
      -- ($(h5fw.east)+(0.97,0)$) -- ++(-0.22,0);
  \begin{scope}[on background layer]
    \node[panel=altblue,
          fit=(h5d)(h5p)(h5w)(h5fd)(h5fw)(matchlbl)] (h5box) {};
  \end{scope}
  \node[htitle, text=altblue!85!black, above=4pt of h5box]
      {H5: Amount Fingerprint};
\end{scope}

\begin{scope}[yshift=\drop cm - 2.75cm]
  \node[dep, minimum width=0.6cm, minimum height=0.36cm] (ld) at (0.3,0){};
  \node[right=3pt of ld, font=\scriptsize]{deposit address};
  \node[wit, minimum width=0.6cm, minimum height=0.36cm] (lw) at (3.7,0){};
  \node[right=3pt of lw, font=\scriptsize]{withdraw address};
  \node[pool] (lp) at (7.4,0){pool};
  \node[right=3pt of lp, font=\scriptsize]{shielded pool};
  \draw[ev, draw=black!50](10.1,0)--++(0.8,0);
  \node[right=3pt, font=\scriptsize] at (10.9,0){linking evidence};
\end{scope}
\end{tikzpicture}
\caption{Illustrating our five heuristics. $\mathbf{H1}$ links deposit and withdraw transactions that belong to the same address. $\mathbf{H2}$ heuristically links deposit and withdraw transactions, if there is at least one on-chain transaction between the deposit and withdraw addresses. $\mathbf{H3}$ links a depositor $d_i$ and withdrawer $w_j$, if the withdraw transaction was sent from the deposit address $d_i$. $\mathbf{H4}$ aims to solve knapsack instances induced by the deposit and withdraw amounts. $\mathbf{H5}$ looks for repeating patterns in the deposit and withdraw amounts that are likely not destroyed by internal or swap transactions. }
\label{fig:heuristic_illustration}
\end{figure}
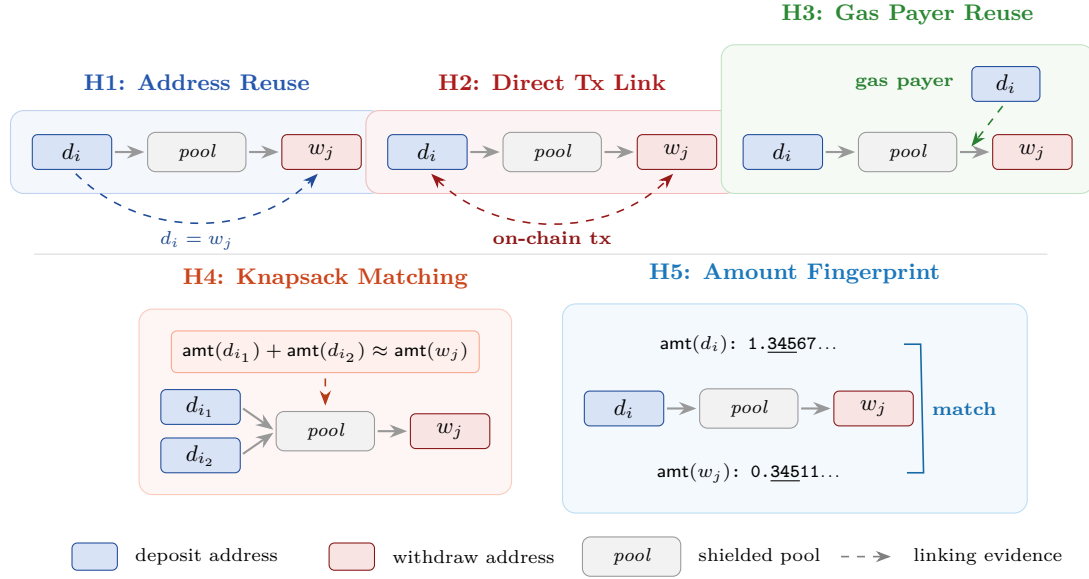
\subsection{Heuristic 4: fee-aware, knapsack matching}\label{sec:heuristic4_knapsack}
Users who withdraw amounts that closely match one or more of their prior deposit amounts reveal a behavioral trace even when there is no direct address or transactional relationship. We formalize this observation as a knapsack-type matching heuristic.

We say that a withdrawal $w_j$ 
\emph{exactly matches} a $k$-to-$\ell$ knapsack pattern with a subset of deposits 
$S = \{d_{i_1},\ldots,d_{i_k}\}$ ($S\subset\mathcal{D}$) if:
\begin{equation}\label{eq:knapsack_equality}
    \sum_{d\in S}\mathsf{amt}(d) = \mathsf{amt}(w_{j_1}) + \cdots + \mathsf{amt}(w_{j_\ell}) \pm \epsilon
\end{equation}
for a small tolerance $\epsilon$ reflecting relayer and other protocol fees.

\begin{definition}(Heuristic 4 (\textbf{H4}))
  A deposit set $S = \{d_{i_1},\ldots,d_{i_k}\}$ and withdrawal set 
  $T = \{w_{j_1},\ldots,w_{j_\ell}\}$ are linked if 
  (i)~all elements of $S$ precede all elements of $T$ in time,
  (ii)~their amounts satisfy the knapsack equality,~\cf~\Cref{eq:knapsack_equality}, within a tolerance $\epsilon$.
\end{definition}

\subsection{Heuristic 5: amount fingerprints}\label{sec:heuristic5_amount_fingerprints}
Unlike Tornado Cash, which enforces fixed deposit denominations, Railgun accepts
arbitrary deposit and withdraw amounts. This flexibility is a privacy liability: users who deposit
a psychologically salient or computationally convenient amount (\eg exactly $1.234\,567$ ETH) tend to preserve the significant digits of that amount in their withdrawal, either through round-tripping or through an internal transfer that preserves the face value minus fees.

We represent the decimal expansion of an amount $a$ as a string of digits 
$\mathsf{dig}(a) = (a_1, a_2, \ldots, a_n)$ at a fixed precision (18 decimal places
for ERC-20 tokens). We define the \emph{fractional fingerprint} of $a$ as 
the three most significant non-zero digits of its fractional part. 
Two amounts $a, b$ are said to be \emph{fingerprint-matched} if their 
fractional fingerprints agree.

\begin{definition}(Heuristic 5 (\textbf{H5}))
  A deposit $d_i$ and withdrawal $w_j$ are heuristically linked by \textbf{H5} if 
  $\mathsf{frac3}({\mathsf{amt}(d_i)}) = \mathsf{frac3}(\mathsf{amt}(w_j))$
  and this fingerprint is globally unique in the dataset,~\ie
  no other deposit-withdrawal pair in the same token shares it.
\end{definition}

\section{Empirical Evaluation of Our Heuristics}\label{sec:evaluation}
We now evaluate the five heuristics introduced in~\Cref{sec:heuristics} against our collected Railgun data set. We first characterize the time-sensitivity of deposit-withdraw pairs, then report the per-heuristic coverage, and finally synthesize the results across all heuristics to estimate the overall anonymity loss,~\cf~\Cref{def:anonymity_measure}. Our evaluation was run on a machine with Intel Core Processor, 8 vCPUs, 16 GB RAM (no swap).
\subsection{Time-sensitivity of withdraw-deposit pairs}\label{sec:time_sensitivity}
A fundamental question for any timing-based analysis is how quickly users tend to withdraw after depositing. Short deposit-withdraw intervals shrink the effective anonymity set dramatically, \eg a withdrawal that occurs one minute after a deposit has only a handful of plausible sources, regardless of how many deposits have accumulated in the pool overall.

\subsubsection{Overall timing distribution}
\Cref{fig:cdf_deltas} shows the empirical distribution of deposit-to-withdraw time differences $\Delta t = t(w_j) - t(d_i)$ for the \num{2904} \textbf{H1}-confirmed pairs. The distribution is strongly right-skewed. The median $\Delta t$ is \num{0.84}~days, meaning that half of all identified users wait less than one day between depositing and withdrawing. The P90 is \num{114.9}~days and the P99 is \num{486.6}~days, reflecting a long tail of users who leave funds in the shielded pool for months. This bimodal character --- fast-cycling users who treat Railgun as a simple mixer, and long-term users who genuinely use the shielded pool --- has direct implications for the effective anonymity set: fast-cycling users expose themselves to timing analysis regardless of the pool's nominal size.
 
\subsubsection{Fee-stratified timing}
We further stratify the \num{2904} \textbf{H1} pairs by the gas price paid on the deposit transaction, splitting them into equal halves at the median gas price (low-fee group: $N = \num{1452}$; high-fee group: $N = \num{1452}$). Counterintuitively, low-gas-price depositors withdraw \emph{faster} than high-gas-price depositors: the low-fee P50 is \num{0.73}~days versus \num{0.98}~days for the high-fee group, and the low-fee P90 is \num{58.9}~days versus \num{178.2}~days. One interpretation is that high-fee depositors are more privacy-aware users who pay for faster block inclusion and then exercise greater patience before withdrawing; low-fee depositors may be less sophisticated users who deposit and withdraw opportunistically. Regardless of the mechanism, both groups have a median withdrawal interval well under one day, making timing a potent signal in practice.

\subsection{Heuristic 1: deposit-withdraw address reuse}\label{sec:heuristic1_evaluation}
\textbf{H1} identifies $\num{2904}$ unshield events (\num{6.10}\% of the filtered unshield population ($\mathcal{W}$) of \num{47596}) as directly reusing a known deposit address, covering \num{8.57}\% of unique addresses and \num{8.52}\% of WETH volume. Every \textbf{H1} link is a certain true positive by construction. We note that \num{710}  ($1.49\%$ of $\vert\mathcal{W}\vert$) deposits were flagged by the PPOI compliance tools, thus, these addresses were forced to withdraw their funds to the very same deposit address. Nevertheless, \textbf{H1} remains our second best performing heuristic in terms of linked withdraw transactions.

\subsection{Heuristic 2: direct transactional linkage}\label{sec:heuristic2_evaluation}
\textbf{H2} identifies \num{4272} unshield events (\num{8.98}\% of $\mathcal{W}$) via at least one on-chain transaction between the deposit and withdraw addresses. The breakdown by asset and direction is shown in~\Cref{fig:h2_edges}: ETH flows from deposit to withdraw account for \num{3139} transactions (all-time), while the reverse direction (withdraw to deposit) accounts for \num{3620} transactions ---the larger of the two, reflecting users who send a portion of their withdrawn funds back to a known deposit address.
ERC20 transfers contribute \num{789} deposit-to-withdraw and \num{1253} withdraw-to-deposit transactions. Restricting to post-2023 activity yields similar proportions, indicating that this behavioral pattern has not decreased as Railgun has matured.
 
The \emph{boundary-metadata} signals --- zero-output unshields and self-broadcast events --- are closely related to \textbf{H2} and together cover \num{6272} (\num{13.18}\%) unshield events. Of these, \num{3433} are flagged by both signals, \num{1996} by the zero-output flag alone, and \num{843} by the self-broadcast flag alone. Self-broadcast events are particularly damaging for privacy: the depositor who acts as their own relayer directly exposes their deposit address in the withdrawal transaction, nullifying the primary anonymity benefit that relayers provide.

\begin{figure}[t!]
\begin{tikzpicture}
  \begin{axis}[
    figurestyle,
    xmode=log, ymode=log,
    xlabel={Distinct recipients served},
    ylabel={Total WETH volume},
    xmin=0.8, ymin=0.0001,
    ymajorgrids, xmajorgrids,
    legend style={at={(0.98,0.02)}, anchor=south east, font=\scriptsize},
  ]
    \addplot+[only marks, mark=triangle*, mark size=2.2pt,
      cteal, opacity=0.65, mark options={draw=cteal, fill=cteal}]
      table [col sep=comma, x=n_unique_recipients, y=total_eth_volume]
      {\TikzAFTDataPath figure_ch4_broadcaster_cluster_points__Mixed-activity.csv};
    \addlegendentry{Mixed-activity (N=1,658)}
    \addplot+[only marks, mark=square*, mark size=2.2pt,
      corange, opacity=0.65, mark options={draw=corange, fill=corange}]
      table [col sep=comma, x=n_unique_recipients, y=total_eth_volume]
      {\TikzAFTDataPath figure_ch4_broadcaster_cluster_points__Self-broadcaster.csv};
    \addlegendentry{Self-broadcaster (N=981)}
    \addplot+[only marks, mark=*, mark size=2.2pt,
      cblue, opacity=0.65, mark options={draw=cblue, fill=cblue}]
      table [col sep=comma, x=n_unique_recipients, y=total_eth_volume]
      {\TikzAFTDataPath figure_ch4_broadcaster_cluster_points__Broadcaster-like.csv};
    \addlegendentry{Broadcasters (N=124)}
  \end{axis}
\end{tikzpicture}
  \caption{Recipient address frequency of withdraw transaction senders. We use these usage statistics to heuristically distinguish between regular users and relayers for withdraw transactions.}
  \label{fig:broadcaster_cluster}
\end{figure}
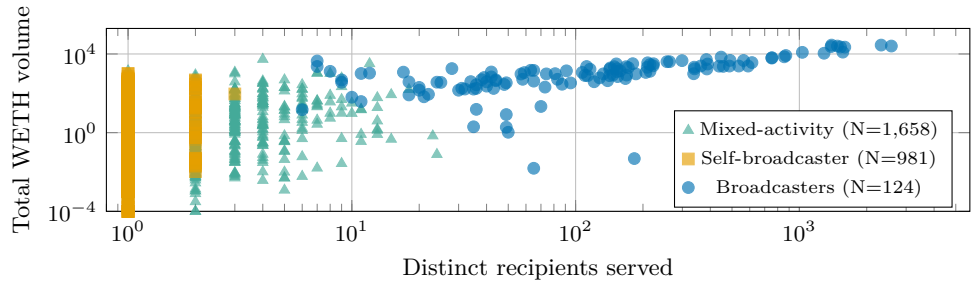

\subsection{Heuristic 3: withdraw transaction gas payer address reuse}\label{sec:heuristic3_evaluation}
We identify $\num{6856}$ withdraw transactions as self-broadcast transactions. These self-broadcast transactions have $\num{2761}$ unique gas payer addresses, and among these there are $\num{1658}$ ($60.1\%$) addresses that are also depositors. There are two mutually exclusive subsets of this withdraw transaction set identified by \textbf{H3}.
\begin{description}
    \item[Gas payer address is the same as $w_i$] If the gas payer address is the same as the withdraw recipient address and this address is also a depositor, then this withdraw transaction has already been linked by \textbf{H1},~\cf~\Cref{sec:heuristic1}. There are $\num{1681}$ withdraw transactions in this subset with $\num{1094}$ distinct gas payer addresses (among these $\num{714}$ were deposit addresses).
    \item[Gas payer address is \emph{not the same} as $w_i$] We found $\num{5175}$ withdraw transactions, that were self-broadcast and the gas payer address was distinct from the withdraw address. More importantly, out of the $\num{1825}$ distinct gas payer addresses in this subset, $\num{1049}$ were also depositor addresses allowing us to link these gas payer/depositor addresses to the withdraw addresses per \textbf{H3}.
\end{description}
\subsection{Heuristic 4: knapsack matching}\label{sec:heuristic4_evaluation}
Although the knapsack problem is an NP-complete problem~\cite{karp2009reducibility}, in certain cases, one has hope to solve the arising real-world knapsack instances efficiently. This is also the case for Railgun. First, the arising knapsack instances are relatively small, and second, the capacities in the knapsack problem are not large. Thus, we chose to apply a pseudopolynomial-time (\ie $\mathcal{O}(nC)$, where $C$ is the knapsack capacity) dynamic programming algorithm to solve our knapsack instances. The algorithm we applied is due to Pisinger~\cite{pisinger2003dynamic}. In our adaptation of his algorithm, we have two parameters $(t,b)$ that determine the running time and the precision of the algorithm's output. In our evaluation, we look primarily for $k$ deposit amounts that sum up to a single withdrawal amount (we call this the $k$-to-$1$ direction). In~\Cref{sec:additional_measurements}, we also evaluate the $1$-to-$k$ direction, \ie where a single deposit amount equals the sum of $k$ different withdraw transactions.

\begin{description}
    \item[Time window $t$] For a given withdraw transaction $w_j$, we only consider deposit transactions $d_i$ for which $t(w_j)-t\leq t(d_i)\leq t(w_j)$. Naturally, as $t$ increases, we need to consider more potential deposit transactions, thus, increasing the algorithm's running time,~\cf~\Cref{table:knapsack_running_time_summary}.
    \item[Bucket size $b$] The dynamic programming algorithm builds a matrix of dimension $n\times C/b$, where $n$ is the number of deposits in our time-window, $C=\mathsf{amt}(w_j)$, and $b$ is the bucket size. Recall our knapsack equation in~\Cref{eq:knapsack_equality}, thus, we have that the error term is $\varepsilon=b/2$. As expected, the larger the bucket is, the faster the algorithm runs (\cf~\Cref{tab:knapsack-runtime}). However, it produces more precise deanonymization results, it incurs higher anonymity loss for Railgun users~\cf~\Cref{fig:knapsack-entropy-histogram-kto1-30d}. 
\end{description}
We observe that running knapsack solver algorithms for reasonable choices of time-window and bucket sizes is completely feasible already on a consumer laptop. We only experienced out-of-memory issues for the all-time and $10^{-5}$ ETH bucket size parameters. We expect that any well-motivated and more funded blockchain analyst (\eg, government-supported agencies, private blockchain analysis companies (Chainalysis\footnote{See:~\url{https://www.chainalysis.com/}}, TRM labs\footnote{See:~\url{https://www.trmlabs.com/}}), etc.) could run various knapsack/SMT solvers on Railgun for $k$-to-$2$ deposits-to-withdraw matchings or perhaps even for $k$-to-$3$ deposits-to-withdraw matchings.
\begin{table}[H]
\centering
\small
\caption{Witness-enumeration runtime per cell, \ie deposit or withdraw, (cap $K=2000$ matching subsets,
witness DFS in C, multi-process pool). Each cell processes all
$|\mathrm{supp}(\mathcal{D})|=29{,}307$ shields ($1$-to-$k$) or
$|\mathrm{supp}(\mathcal{W})|=38{,}472$ unshields ($k$-to-$1$). The
all-time $10^{-5}$ ETH bucket is skipped on both sides: it would require
$\sim 50$\,GiB per worker that is not supported by our limited experimental setup.}
\label{tab:knapsack-runtime}
\begin{tabular}{@{}lrrrr@{}}
\toprule
& \multicolumn{4}{c}{Bucket size ($b$ ETH)} \\
\cmidrule(l){2-5}
Time window ($t$ days) & $10^{-2}$ & $10^{-3}$ & $10^{-4}$ & $10^{-5}$ \\
\midrule
\multicolumn{5}{l}{\textbf{(a) $1$-to-$k$ direction} (laptop, Intel Core Ultra 7 155U, 19\,GiB WSL RAM)} \\
\midrule
3\,days        & 0.9\,s   & 2.1\,s   & 6.8\,s             & 44.7\,s \\
7\,days        & 2.5\,s   & 3.5\,s   & 10.1\,s            & 1\,min 14\,s \\
30\,days       & 4.7\,s   & 10.5\,s  & 36.3\,s            & 6\,min 36\,s \\
180\,days      & 22.4\,s  & 56.7\,s  & 4\,min 41\,s       & 1\,h 12\,min 45\,s \\
all-time       & 1\,min 15\,s & 2\,min 34\,s & 15\,min 41\,s & ---\,(skip) \\
\midrule
\multicolumn{5}{l}{\textbf{(b) $k$-to-$1$ direction} (laptop, Intel Core Ultra 7 155U, 19\,GiB WSL RAM)} \\
\midrule
3\,days        & 0.9\,s   & 1.2\,s   & 2.4\,s             & 21.6\,s \\
7\,days        & 1.2\,s   & 1.9\,s   & 4.7\,s             & 49.0\,s \\
30\,days       & 2.4\,s   & 4.6\,s   & 20.7\,s            & 4\,min 56\,s \\
180\,days      & 12.0\,s  & 20.1\,s  & 1\,min 50\,s       & 34\,min 24\,s \\
all-time       & 24.8\,s  & 57.1\,s  & 7\,min 46\,s       & ---\,(skip) \\
\bottomrule
\end{tabular}
\label{table:knapsack_running_time_summary}
\end{table}

In~\Cref{fig:knapsack-entropy-heatmap-kto1-30d} we study the capabilities of knapsack solver algorithms to decrease the anonymity guarantees of Railgun. For each withdraw $w_j$, we take the optimistic anonymity measure as the Shannon entropy of the distribution that assigns equal probabilities to all prior deposit transactions,~\cf~\Cref{def:naive_anonymity}. In contrast, we applied our adaptation of Pisinger's algorithm to $\mathsf{amt}(w_j)$. Typically, the algorithm outputs several candidate solutions for a given withdraw amount $\mathsf{amt}(w_j)$. We count how many times each deposit $d_i$ appears in the candidate solutions produced by Pisinger's algorithm. We convert these frequencies into a probability distribution and take its Shannon entropy to measure the reduced anonymity guarantees provided by our knapsack solver algorithm. For instance, in~\Cref{fig:knapsack-entropy-heatmap-kto1-30d}, we find that for the parameters $(t,b)=(30\ \textrm{days},10^{-5}\ \textrm{ETH})$, the median anonymity loss for withdrawals is $3.42$ bits of entropy. For this parameter set, the knapsack solver algorithm has found $85$ withdraw transactions to which there is a single subset of deposit transactions that sum up exactly to $\mathsf{amt}(w_j)$.

\subsection{Heuristic 5: Amount Fingerprints}\label{sec:heuristic5_evaluation}

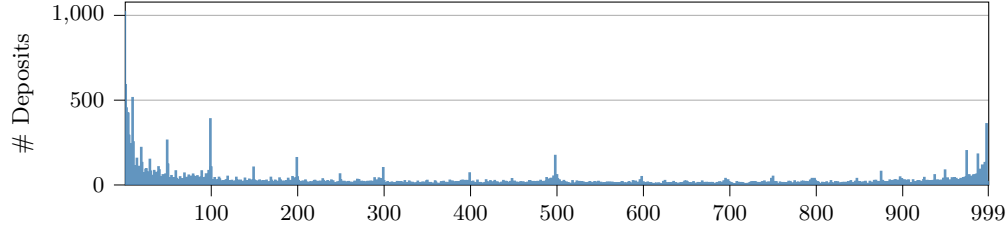
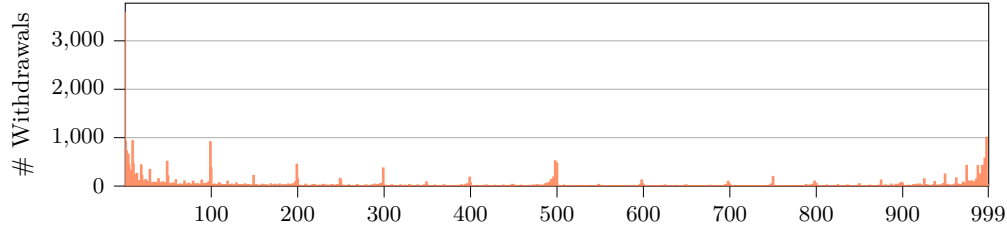
\begin{figure}[t!]

\subfloat[Deposit amounts grouped by first 3 fractional digits (000--999)]{%
\begin{tikzpicture}
\begin{axis}[
    figurestyle,
    ybar,
    bar width=1pt,
    tick align=outside,
    tick pos=left,
    xmin=0.5, xmax=999.5,
    ymin=0, ymax=1078.35,
    xtick={0,100,200,300,400,500,600,700,800,900,999},
    x grid style={darkgrey176},
    y grid style={darkgrey176},
    ymajorgrids,
    xtick style={color=black},
    ytick style={color=black},
    ylabel={\# Deposits},
]
\addplot+[draw=none, fill=steelblue, fill opacity=0.85]
    table[col sep=comma, x=bin, y=deposits_count]
    {Figures/frac3_fingerprint_histogram_data.csv};
\end{axis}
\end{tikzpicture}%
\label{fig:frac3-deposit-histogram}
}

\par\medskip

\subfloat[Withdrawal amounts grouped by first 3 fractional digits (000--999)]{%
\begin{tikzpicture}
\begin{axis}[
    figurestyle,
    ybar,
    bar width=1pt,
    tick align=outside,
    tick pos=left,
    xmin=0.5, xmax=999.5,
    ymin=0, ymax=3777.9,
    xtick={0,100,200,300,400,500,600,700,800,900,999},
    x grid style={darkgrey176},
    y grid style={darkgrey176},
    ymajorgrids,
    xtick style={color=black},
    ytick style={color=black},
    ylabel={\# Withdrawals},
]
\addplot+[draw=none, fill=coral, fill opacity=0.85]
    table[col sep=comma, x=bin, y=withdrawals_count]
    {Figures/frac3_fingerprint_histogram_data.csv};
\end{axis}
\end{tikzpicture}%
\label{fig:frac3-withdrawal-histogram}
}
\caption{Distribution of deposit and withdrawal amounts by the first three fractional digits.}
  \label{fig:frac3_fingerprint_histogram}
\end{figure}

Finally, we examine whether deposit and withdraw transaction amounts themselves leave fingerprints usable for deanonymization. 
In particular, we look at the distribution of the first three digits after the decimal 
point for both deposit and withdrawal amounts~\cf~\Cref{fig:frac3_fingerprint_histogram}.

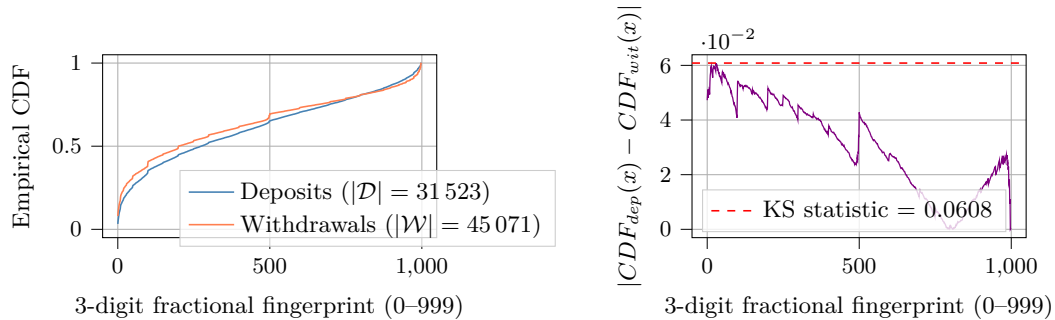
\begin{figure}[ht!]

\subfloat[Empirical cumulative distribution functions (CDF)]{%
\begin{tikzpicture}
\begin{axis}[
figurestyle,halfwidth,
legend cell align={left},
legend style={
  fill opacity=0.8,
  draw opacity=1,
  text opacity=1,
  at={(0.23,0.37)},
  anchor=north west,
  draw=lightgrey204
},
tick align=outside,
tick pos=left,
x grid style={darkgrey176},
xlabel={3-digit fractional fingerprint (0–999)},
xmajorgrids,
xmin=-49.95, xmax=1048.95,
xtick style={color=black},
y grid style={darkgrey176},
ylabel={Empirical CDF},
ymajorgrids,
ymin=-0.0499767034234874, ymax=1.04999889063921,
ytick style={color=black}
]
\addplot [semithick, steelblue]
table [x=x, y=y, col sep=comma] {Figures/frac3_ks_cdf_deposits.csv};
\addlegendentry{Deposits ($\vert\mathcal{D}\vert=\num{31523}$)}
\addplot [semithick, coral]
table [x=x, y=y, col sep=comma] {Figures/frac3_ks_cdf_withdrawals.csv};
\addlegendentry{Withdrawals ($\vert\mathcal{W}\vert=\num{45071}$)}
\end{axis}
\end{tikzpicture}%
}
\hfill
\subfloat[CDF difference]{%
\begin{tikzpicture}
\begin{axis}[
figurestyle,halfwidth,
legend cell align={left},
legend style={
  fill opacity=0.8,
  draw opacity=1,
  text opacity=1,
  at={(0.03,0.03)},
  anchor=south west,
  draw=lightgrey204
},
tick align=outside,
tick pos=left,
x grid style={darkgrey176},
xlabel={3-digit fractional fingerprint (0–999)},
xmajorgrids,
xmin=-49.95, xmax=1048.95,
xtick style={color=black},
y grid style={darkgrey176},
ylabel={$|CDF_{dep}(x) - CDF_{wit}(x)|$},
ymajorgrids,
ymin=-0.00304205210502105, ymax=0.063883094205442,
ytick style={color=black}
]
\addplot [line width=0.48pt, purple, forget plot]
table [x=x, y=y, col sep=comma] {Figures/frac3_ks_cdf_difference.csv};
\addplot [semithick, red, dashed]
table [x=x, y=y, col sep=comma] {Figures/frac3_ks_stat_line.csv};
\addlegendentry{KS statistic = 0.0608}
\end{axis}
\end{tikzpicture}%
}
  \caption{Kolmogorov--Smirnov test (KS-stat=$0.0608$, $p=3.95 \times 10^{-60}$) for fractional-digit deposit and withdraw amount fingerprints,~\cf~\Cref{fig:frac3_fingerprint_histogram}. The KS-test rejects the null-hypothesis.}
  \label{fig:frac3_ks_test}
\end{figure}

As the histogram in~\Cref{fig:frac3_fingerprint_histogram} shows, the distribution 
of fractional fingerprints is highly skewed: a small number of round-number fingerprints 
(\eg all zeros) are extremely common, while the majority of fingerprints appear only a 
handful of times. This confirms that deposit and withdrawal amounts are not random 
decimal strings. We also experimented with using such amount fingerprints for 
deanonymization. However, large-value patterns are already largely captured by the 
knapsack heuristic, while the least significant digits of small amounts are likely 
obscured by transaction costs and gas-price choices, which are not observable from the 
shielded-pool amounts alone. We therefore did not find fractional amount fingerprints 
sufficiently reliable for deanonymization. Nevertheless, 
\Cref{fig:frac3_fingerprint_histogram} provides a useful diagnostic view of the empirical 
distribution of the first three fractional digits.

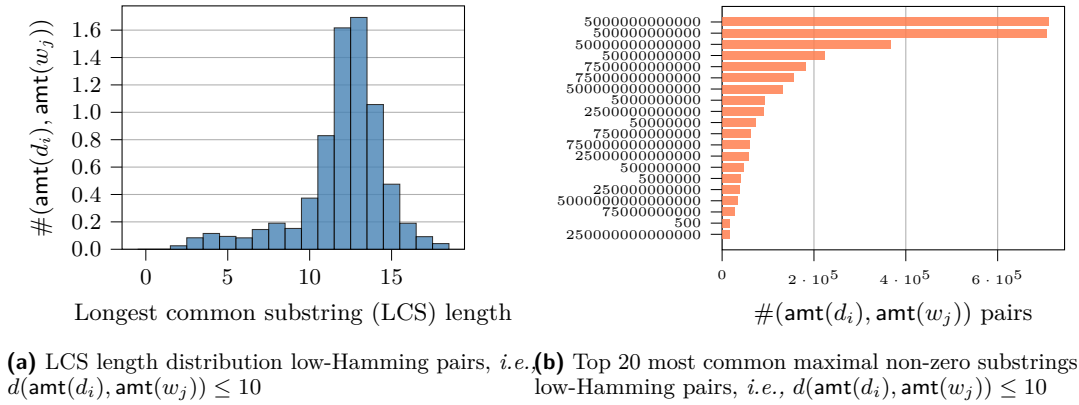
\begin{figure}[h]
\subfloat[LCS length distribution {low-Hamming pairs, \ie $d(\mathsf{amt}(d_i),\mathsf{amt}(w_j)) \leq 10$}]{
\begin{tikzpicture}
\begin{axis}[
  figurestyle,
   width=0.47*\figurewidth,
   height=1.2*\figureheight,
  tick align=outside,
  tick pos=left,
  x grid style={darkgrey176},
  xlabel={Longest common substring (LCS) length},
  xmin=-1.45, xmax=19.45,
  xtick style={color=black},
  y grid style={darkgrey176},
  ylabel={\#($\mathsf{amt}(d_i),\mathsf{amt}(w_j)$)},
  ymajorgrids,
  ymin=0, ymax=1776976.95,
  ytick style={color=black},
  scaled y ticks=false,
  ytick={0,200000,400000,600000,800000,1000000,1200000,1400000,1600000,1800000},
  yticklabels={0.0,0.2,0.4,0.6,0.8,1.0,1.2,1.4,1.6,1.8},
]
\draw[draw=black,fill=steelblue,opacity=0.8] (axis cs:-0.5,0) rectangle (axis cs:0.5,0);
\draw[draw=black,fill=steelblue,opacity=0.8] (axis cs:0.5,0) rectangle (axis cs:1.5,0);
\draw[draw=black,fill=steelblue,opacity=0.8] (axis cs:1.5,0) rectangle (axis cs:2.5,25370);
\draw[draw=black,fill=steelblue,opacity=0.8] (axis cs:2.5,0) rectangle (axis cs:3.5,83790);
\draw[draw=black,fill=steelblue,opacity=0.8] (axis cs:3.5,0) rectangle (axis cs:4.5,115299);
\draw[draw=black,fill=steelblue,opacity=0.8] (axis cs:4.5,0) rectangle (axis cs:5.5,93784);
\draw[draw=black,fill=steelblue,opacity=0.8] (axis cs:5.5,0) rectangle (axis cs:6.5,83291);
\draw[draw=black,fill=steelblue,opacity=0.8] (axis cs:6.5,0) rectangle (axis cs:7.5,144272);
\draw[draw=black,fill=steelblue,opacity=0.8] (axis cs:7.5,0) rectangle (axis cs:8.5,190382);
\draw[draw=black,fill=steelblue,opacity=0.8] (axis cs:8.5,0) rectangle (axis cs:9.5,151972);
\draw[draw=black,fill=steelblue,opacity=0.8] (axis cs:9.5,0) rectangle (axis cs:10.5,373329);
\draw[draw=black,fill=steelblue,opacity=0.8] (axis cs:10.5,0) rectangle (axis cs:11.5,829475);
\draw[draw=black,fill=steelblue,opacity=0.8] (axis cs:11.5,0) rectangle (axis cs:12.5,1616322);
\draw[draw=black,fill=steelblue,opacity=0.8] (axis cs:12.5,0) rectangle (axis cs:13.5,1692359);
\draw[draw=black,fill=steelblue,opacity=0.8] (axis cs:13.5,0) rectangle (axis cs:14.5,1056763);
\draw[draw=black,fill=steelblue,opacity=0.8] (axis cs:14.5,0) rectangle (axis cs:15.5,474919);
\draw[draw=black,fill=steelblue,opacity=0.8] (axis cs:15.5,0) rectangle (axis cs:16.5,189874);
\draw[draw=black,fill=steelblue,opacity=0.8] (axis cs:16.5,0) rectangle (axis cs:17.5,92072);
\draw[draw=black,fill=steelblue,opacity=0.8] (axis cs:17.5,0) rectangle (axis cs:18.5,41475);
\end{axis}
\end{tikzpicture}
\label{fig:lcs-length-distribution}
}
\hfill
\subfloat[Top 20 most common maximal non-zero substrings
  {low-Hamming pairs}, \ie $d(\mathsf{amt}(d_i),\mathsf{amt}(w_j)) \leq 10$]{
\begin{tikzpicture}
\begin{axis}[
  figurestyle,
  width=0.47*\figurewidth,
   height=1.2*\figureheight,
  tick align=outside,
  tick pos=left,
   x grid style={darkgrey176},
  xlabel={\#($\mathsf{amt}(d_i),\mathsf{amt}(w_j)$) pairs},
  xmajorgrids,
  xmin=0, xmax=745693.2,
  xtick style={color=black},
  y dir=reverse,
  y grid style={darkgrey176},
  ymin=-1.39, ymax=20.39,
  ytick style={color=black},
  scaled x ticks=false,
  ytick={0,1,2,3,4,5,6,7,8,9,10,11,12,13,14,15,16,17,18,19},
  yticklabels={
    5000000000000,
    500000000000,
    50000000000000,
    50000000000,
    7500000000000,
    75000000000000,
    500000000000000,
    5000000000,
    2500000000000,
    50000000,
    750000000000,
    750000000000000,
    25000000000000,
    500000000,
    5000000,
    250000000000,
    5000000000000000,
    75000000000,
    500,
    250000000000000
  },
   tick label style={font=\tiny},
]
\draw[draw=none,fill=coral,fill opacity=0.85] (axis cs:0,-0.4) rectangle (axis cs:710184,0.4);
\draw[draw=none,fill=coral,fill opacity=0.85] (axis cs:0,0.6) rectangle (axis cs:705604,1.4);
\draw[draw=none,fill=coral,fill opacity=0.85] (axis cs:0,1.6) rectangle (axis cs:368001,2.4);
\draw[draw=none,fill=coral,fill opacity=0.85] (axis cs:0,2.6) rectangle (axis cs:224167,3.4);
\draw[draw=none,fill=coral,fill opacity=0.85] (axis cs:0,3.6) rectangle (axis cs:181652,4.4);
\draw[draw=none,fill=coral,fill opacity=0.85] (axis cs:0,4.6) rectangle (axis cs:156044,5.4);
\draw[draw=none,fill=coral,fill opacity=0.85] (axis cs:0,5.6) rectangle (axis cs:130857,6.4);
\draw[draw=none,fill=coral,fill opacity=0.85] (axis cs:0,6.6) rectangle (axis cs:92128,7.4);
\draw[draw=none,fill=coral,fill opacity=0.85] (axis cs:0,7.6) rectangle (axis cs:89739,8.4);
\draw[draw=none,fill=coral,fill opacity=0.85] (axis cs:0,8.6) rectangle (axis cs:71949,9.4);
\draw[draw=none,fill=coral,fill opacity=0.85] (axis cs:0,9.6) rectangle (axis cs:62452,10.4);
\draw[draw=none,fill=coral,fill opacity=0.85] (axis cs:0,10.6) rectangle (axis cs:59710,11.4);
\draw[draw=none,fill=coral,fill opacity=0.85] (axis cs:0,11.6) rectangle (axis cs:57688,12.4);
\draw[draw=none,fill=coral,fill opacity=0.85] (axis cs:0,12.6) rectangle (axis cs:46384,13.4);
\draw[draw=none,fill=coral,fill opacity=0.85] (axis cs:0,13.6) rectangle (axis cs:39859,14.4);
\draw[draw=none,fill=coral,fill opacity=0.85] (axis cs:0,14.6) rectangle (axis cs:38374,15.4);
\draw[draw=none,fill=coral,fill opacity=0.85] (axis cs:0,15.6) rectangle (axis cs:33588,16.4);
\draw[draw=none,fill=coral,fill opacity=0.85] (axis cs:0,16.6) rectangle (axis cs:28215,17.4);
\draw[draw=none,fill=coral,fill opacity=0.85] (axis cs:0,17.6) rectangle (axis cs:17095,18.4);
\draw[draw=none,fill=coral,fill opacity=0.85] (axis cs:0,18.6) rectangle (axis cs:16154,19.4);
\end{axis}
\end{tikzpicture}
\label{fig:common-substrings}
}
\caption{Longest common subsequence length distribution (left) and the most common maximal non-zero substrings among low-Hamming deposit-withdraw transaction amount pairs (right).}
\label{fig:lcs-substrings}
\end{figure}

To further inspect this effect, we compare deposit--withdraw transaction amount pairs with small
Hamming distance as decimal strings. \Cref{fig:lcs-substrings} shows that these
pairs often share long common substrings (typically 11--14 digits), and that the
most frequent maximal non-zero substrings are dominated by round-denomination
patterns (\eg long runs of zeros after a small leading digit). Overall, amount
strings exhibit structure, but it is driven mainly by global denomination/rounding
effects rather than stable user-specific fingerprints. Consequently, beyond the
knapsack signal, the remaining low-order digits are too noisy for reliable
standalone deanonymization.

We further tested whether these fractional-digit patterns provide a robust signal.
The Kolmogorov--Smirnov test in~\Cref{fig:frac3_ks_test} indicates that the empirical
distributions of deposit and withdrawal transaction amount fingerprints are statistically distinguishable.
However, this difference is mainly driven by global rounding and denomination effects,
rather than by stable pair-specific fingerprints.

The Hamming-distance diagnostics in~\Cref{fig:hamming_distance_histogram}
--\ref{fig:unique_fingerprint_substring_lengths} support the same conclusion.
Low-distance deposit--withdraw pairs do exist, and their differences are concentrated
towards later fractional positions, while early fractional digits often agree. Moreover,
unique common substrings are typically short, with most value-fingerprint substrings
appearing at lengths around four to six digits. Overall, these patterns confirm that
amount strings contain structure, but we did not find this structure sufficiently
specific or stable to serve as an independent deanonymization heuristic beyond the
knapsack-based analysis.

\subsection{Summary of Our Results}\label{sec:summary}
We provide an overview of all our heuristic findings in~\Cref{tab:summary_table}. We observe that \textbf{H2} produces the largest number of heuristic deposit-withdraw links. Altogether, we find that the first four heuristics link $17.65\%$ of all Ethereum L1 Railgun withdraws. Note that these are only the one-to-one links ($d_i\leftrightarrow w_j$) output by our heuristics. We study the probabilistic (\ie non-unique) deposit-withdraw links and anonymity loss incurred by the knapsack matching Heuristic in~\Cref{fig:knapsack-entropy-heatmap-kto1-30d,fig:knapsack-entropy-histogram-kto1-30d}.

\begin{table}[t!]
\centering
\small
\begin{tabular}{@{}lr@{}}
\toprule
\textbf{Set} & \textbf{Cardinality} \\
\midrule
$\mathcal{W}$ & \num{47596} \\
$\mathcal{D}$ & \num{34747} \\
$\mathrm{supp}(\mathcal{W})$ & \num{38472} \\
$\mathrm{supp}(\mathcal{D})$ & \num{29307} \\
\textbf{H1} (address reuse)                            & \num{2904} ($6.10\%$) \\
\quad of which PPOI-flagged                                    & \num{710}  ($1.49\%$) \\
\textbf{H2} (direct tx link)               & \num{4272} ($8.98\%$) \\
\textbf{H1}$\cup$\textbf{H2}                                   & \num{6003} ($12.61\%$) \\
\textbf{H3} (gas payer reuse)                          & \num{1049} ($2.20\%$) \\
\textbf{H1}$\cup$\textbf{H2}$\cup$\textbf{H3}                  & \num{6715} ($14.11\%$) \\
\textbf{H4} (knapsack matching)           & \num{2097} ($4.41\%$) \\
\midrule
\textbf{Total} (\textbf{H1}$\cup$\textbf{H2}$\cup$\textbf{H3}$\cup$\textbf{H4}) & \num{8398} ($17.65\%$) \\
\bottomrule
\end{tabular}
\caption{Summary of our findings. Cardinalities are counted at the raw unshield event level (one entry per filtered unshield in $\mathcal{W}$); percentages are relative to $|\mathcal{W}| = 47{,}596$. \textbf{H1}--\textbf{H3} apply the event-level causal rule from~\Cref{sec:heuristics}; \textbf{H4} is the union over all $5\times 4$ knapsack cells of the $k$-to-$1$ witness pass of unshield super-rows whose matching subsets implicate a single depositor address ($H_{\mathrm{knap,addr}}=0$), expanded to raw events via $\texttt{w\_agg\_size}$.}
\label{tab:summary_table}
\end{table}

\subsection{Privacy Recommendations}\label{sec:privacy_recommendations}
Our findings point directly to actionable recommendations for users who wish to obtain stronger privacy guarantees from Railgun. We organize these by the heuristic they address.
 
\myparagraph{Always use a relayer (H3).}
The single most impactful recommendation is to always use a third-party relayer when issuing withdraw transactions. Among the \num{6856} self-broadcast withdraw transactions in our dataset, \num{5175} were sent from an address distinct from the withdraw recipient, and \num{1049} of those gas-payer addresses were also depositor addresses, making them immediately linkable via \textbf{H3}. A relayer's address is independent of the user's deposit and withdraw addresses, eliminating this leakage channel entirely. Our broadcaster cluster analysis identified \num{124} active relayer-like addresses that together serve \num{89.12}\% of all relayed volume, so relayer availability is not a practical obstacle. Users should prefer relayers that have served a large number of distinct recipients, as these provide larger anonymity sets.
 
\myparagraph{Never reuse deposit and withdraw addresses (H1).}
\textbf{H1} is the only heuristic that produces zero false positives by construction:
if the same Ethereum address appears on both sides of the shielded pool, the link is certain. Yet \num{2904} (\num{6.10}\%) of unshield events in our dataset reuse a known deposit address. Users should generate a fresh Ethereum address for every withdraw transaction and should never fund that address from a known deposit address prior to withdrawing.
 
\myparagraph{Avoid direct on-chain transactions between deposit and withdraw
addresses (H2).}
\textbf{H2} identifies \num{4272} (\num{8.98}\%) of unshield events via at least one on-chain transaction between the deposit and withdraw address.
The most common pattern is users topping up a withdraw address with ETH from their deposit address to cover gas fees --- a problem that is fully solved by using a relayer (see above). Users should treat the deposit address and the withdraw address as completely disjoint identities with no on-chain contact between them, not even through intermediary contracts or decentralized exchanges.
 
\myparagraph{Wait longer before withdrawing (timing).}
The median deposit-to-withdraw interval for \textbf{H1}-identified pairs is only \num{0.84}~days. A withdrawal that occurs minutes or hours after a deposit has a very small effective anonymity set, regardless of the pool's nominal size. Users who are genuinely concerned about their financial privacy should leave funds in the shielded pool for weeks or months, during which the pool accumulates additional deposits that expand their anonymity set. Our Little's Law estimate suggests a characteristic pool residence time of roughly \num{183}~days for the average deposited ETH; users who withdraw much faster than this are outliers and more easily identified.
 
\myparagraph{Avoid psychologically salient or round amounts (H4, H5).}
\textbf{H4} exploits knapsack relationships between deposit and withdraw amounts. \textbf{H5} exploits the fact that users tend to preserve distinctive digit patterns across transactions. Both heuristics are rendered ineffective if the amounts that enter and leave the shielded pool bear no predictable relationship to each other. In practice, this means: avoid withdrawing exactly the amount deposited; avoid splitting or merging deposits in simple integer ratios; and avoid depositing amounts with highly distinctive fractional digits (\eg \texttt{1.234\,567} ETH) that are unlikely to be destroyed by internal transactions or swap fees. Ideally, users should perform at least one internal transaction or private swap between depositing and withdrawing, so that the amount that ultimately leaves the pool bears no simple arithmetic relationship to the deposited amount.
 
\myparagraph{Use the full functionality of the shielded pool.}
We saw that \num{15.00}\% of deposited ETH remains in the shielded pool as of our measurement date, and that internal transactions are actively used (\num{17767} internal transfers over the observation period). Users who perform internal transactions and private swaps before withdrawing are more resistant to \textbf{H4} and \textbf{H5}, since these operations transform the amount in ways that break simple arithmetic linkages. Importantly, users who leave funds in the pool for longer periods contribute to a larger anonymity set for all participants. Privacy in a shielded pool is a public good, and short-cycling users who deposit and immediately withdraw not only expose themselves but also reduce the privacy of other users
by shrinking the effective anonymity set.
 
\myparagraph{Protocol-level recommendations.}
Beyond individual user behavior, our findings suggest several design-level
improvements that the Railgun protocol could adopt to reduce behavioral leakage. First, the protocol could enforce a minimum residence time or introduce time-locked notes, making rapid deposit-withdraw cycling impossible. Second, Railgun could follow Tornado Cash's model of offering fixed denomination tiers as an option, which would eliminate \textbf{H4} and \textbf{H5} for users who opt into that mode. Third, the relayer infrastructure could be improved to make relayer selection more uniform across the broadcaster population, reducing the fingerprinting potential of relayer choice patterns. We leave the formal analysis of the privacy improvements achievable by such protocol modifications to future work.
\section{Related Work}\label{sec:related_work}
Cryptocurrency transaction graph analysis has extensive literature~\cite{meiklejohn2013fistful,ron2013quantitative,victor2019measuring,victor2021wash}. Many works apply machine learning tools in the context of enforcing KYC/AML regulations or extracting intelligence from public transaction graphs~\cite{pham2016anomaly,harlev2018breaking,weber2019aml}. Next, we focus only on the literature on the deanonymization of privacy-enhancing tools on top of various blockchains.
\subsection{Deanonymization of Bitcoin mixers}
The earliest work on breaking unlinkability in Bitcoin focused on CoinJoin~\cite{maxwell2013coinjoin,stockinger2021pinpointing}, which merges the inputs of multiple users into a single transaction~\cite{gavenda2025analysis}. Möser et al.~\cite{moser2017anonymous} showed that the small anonymity sets typical of deployed CoinJoin implementations allow a substantial fraction of transactions to be 
deanonymized. Later work identified address reuse, amount clustering, and transaction graph structure as dominant leakage channels~\cite{biryukov2014deanonymisation,goldfeder2018cookie}. More recent large-scale empirical analyses of decentralized CoinJoin services, including Wasabi and Samourai, showed that even when the mixing transaction itself provides a larger nominal anonymity set, privacy can be substantially reduced by pre- and post-mixing behavior~\cite{svenda2026ecosystem}, address-selection patterns, and subsequent flows to exchanges~\cite{stutz2022adoption}.

\subsection{Tornado Cash}
The closest prior work to ours is the empirical analysis of Tornado Cash 
by Béres et al.~\cite{beres2021blockchain}, which introduced several of the 
heuristics we adapt and extend, including address reuse, direct transactional 
linkage, and value fingerprinting via the Danaan-gift attack. Wu et al.\ developed 
Tutela~\cite{wu2022tutela}, an open-source tool that operationalizes these heuristics. 
Wang et al.~\cite{wang2023zero} studied how zero-knowledge proof mixers improve 
and worsen user privacy, finding that optional shielding creates self-selection biases 
that reduce the effective anonymity set. Tang et al.~\cite{tang2021analysis} analyzed 
address linkability in Tornado Cash specifically. Our work differs in focusing on 
Railgun, which offers richer in-pool functionality and supports arbitrary token amounts, 
creating new leakage channels not present in fixed-denomination mixers.

Complementary to deanonymization work, Wang et al.~\cite{wang2023payless} studied the cost side of on-chain mixers, showing how Tornado-Cash-like privacy pools can be made more cost-effective. While their focus is mixer design and transaction cost, our work studies the residual privacy loss that arises from user behavior and amount-level leakage in a richer shielded-pool setting.

\subsection{Monero, Zcash}
Möser et al.~\cite{moser2018empirical} provided a large-scale empirical analysis 
of traceability in Monero, showing that the zero-mixin era and subsequent 
hard forks introduced significant traceability. Temporal correlation attacks 
on ring member selection have been studied analytically, showing that 
non-uniform ring sampling degrades anonymity substantially~\cite{yu2019ringattack}. 
These results parallel our findings: the practical anonymity set is often smaller than the nominal cryptographic set.

Ye et al.~\cite{ye2020altcoin} studied traceability in Zcash's shielded pool, 
finding that founder and miner withdrawal behavior follows recognizable periodic 
patterns that reduce the effective anonymity of their transactions. 
The general lesson --- that behavioral regularity defeats cryptographic privacy~\cite{biryukov2018deanonymization,biryukov2019privacy,kappos2018empirical} --- 
directly motivates our work.

\subsection{Network-layer deanonymization}
Several works have studied privacy attacks at the network layer rather than 
the blockchain layer. Biryukov et al.~\cite{biryukov2014deanonymisation} exploited P2P connection patterns to link IP addresses to Bitcoin addresses. Heimbach et al.~\cite{heimbach2025deanonymizing} applied similar techniques to Ethereum. Wang et al.~\cite{wang2025time} demonstrated a timing-correlation attack against RPC users achieving over 95\% success rate. Our work is complementary: we restrict ourselves to on-chain data, and our heuristics could be combined with network-layer observations for 
stronger deanonymization.

\subsection{Wallet and gas fingerprinting}
Soleti et al.~\cite{soleti2025attacking} demonstrated that wallet software of Ethereum users can  be identified from gas-pricing strategies, providing a fingerprint orthogonal  to the amount and timing features we study. We leave the combination of wallet fingerprints with our heuristics to future work.
\section{Conclusion and Future Directions}\label{sec:conclusion}
This work quantitatively assessed the practical privacy guarantees of Railgun users. We found that $17.65\%$ of Railgun withdraw transactions can be heuristically linked to \emph{a single deposit} transaction. However, we believe that in reality, the actual achieved privacy guarantees are even weaker due to the multiple layers (network, application, etc.) and heuristics we have not studied in this paper. We foresee the following fruitful directions for future work.
\begin{description}
    \item[Wallet fingerprints] We did not exploit many types of metadata, i.e., typically it is possible to assess which wallet software an entity uses. For instance, different wallets use various algorithms to compute their applied transaction fees. These wallet gas fingerprints~\cite{soleti2025attacking} could be used to further decrease the empirical Railgun anonymity set.  
    \item[Graph-embedding tools] We did not apply machine learning tools to learn the Ethereum transaction graph~\cite{beres2021blockchain} and use it to link deposit and withdraw addresses. 
    \item[Multi-layered privacy analysis] So far, this paper has only focused on the Ethereum blockchain data. A more resourceful privacy analysis should take full advantage of network~\cite{heimbach2025deanonymizing}, application~\cite{torres2023your}, cross-chain and other leakages as well. One may even consider ``active'' privacy attacks, \eg when the adversary is one of the Railgun relayers or users.  
\end{description}

\ifanonymous
\else
\myparagraph{Acknowledgements}
István András Seres was supported by the Ministry of Culture and Innovation and the National Research, Development, and Innovation Office (NKFIH) within the Quantum Information National Laboratory of Hungary (Grant No. 2022-2.1.1-NL-2022-00004).
János Tapolcai was supported by Grant No.  K\_23 146347 of NKFIH.
\fi
\bibliographystyle{plainurl}
\bibliography{references}

\appendix
\section{Additional Measurements}\label{sec:additional_measurements}
Due to space constraints, we enclose many of our measurements in this section.


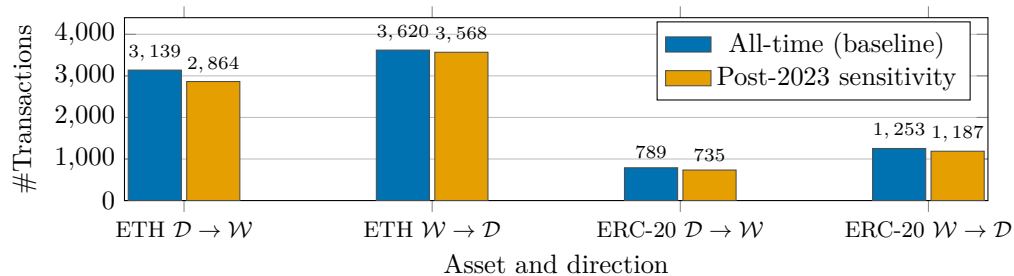
\begin{figure}
\begin{tikzpicture}
  \begin{axis}[
    width=\figurewidth, height=\figureheight,
    ybar, /pgf/bar width=20pt, enlarge x limits=0.08, ymin=0, ymax=4400, area legend,
    xtick={0,1,2,3},
    xticklabels={{ETH $\mathcal{D}\rightarrow\mathcal{W}$},{ETH $\mathcal{W}\rightarrow\mathcal{D}$},{ERC-20 $\mathcal{D}\rightarrow\mathcal{W}$},{ERC-20 $\mathcal{W}\rightarrow\mathcal{D}$}},
    x tick label style={align=center, font=\footnotesize},
    xlabel={Asset and direction},
    ylabel={$\#$Transactions}, ymajorgrids,
    legend style={at={(0.98,0.98)}, anchor=north east},
    nodes near coords,
    every node near coord/.append style={
      font=\scriptsize,
      /pgf/number format/.cd, fixed, precision=0, 1000 sep={,}, /tikz/.cd},
  ]
    \addplot[fill=cblue, draw=black!70] coordinates {(0,3139) (1,3620) (2,789) (3,1253)};
    \addlegendentry{All-time (baseline)}
    \addplot[fill=corange, draw=black!70] coordinates {(0,2864) (1,3568) (2,735) (3,1187)};
    \addlegendentry{Post-2023 sensitivity}
  \end{axis}
\end{tikzpicture}
\caption{The prevalence of transactions  across depositor ($\mathcal{D}$) and withdrawer ($\mathcal{W}$) addresses motivates the formulation of Heuristic 2,~\cf~\Cref{sec:heuristic2}.}
\label{fig:h2_edges}
\end{figure}

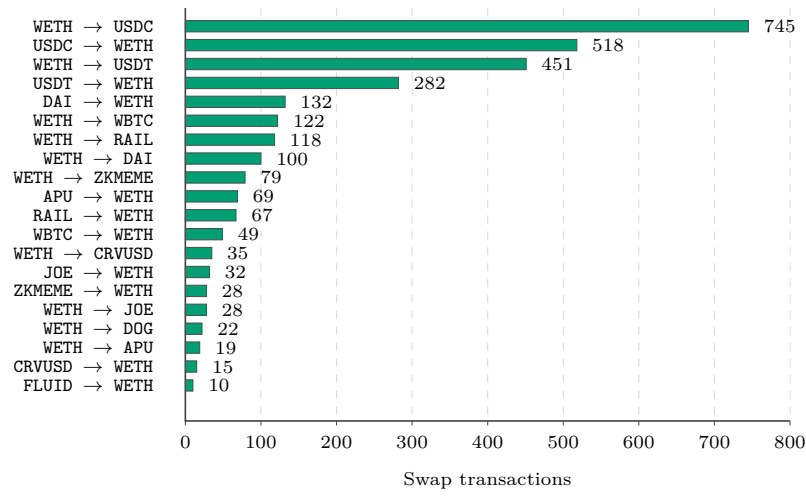
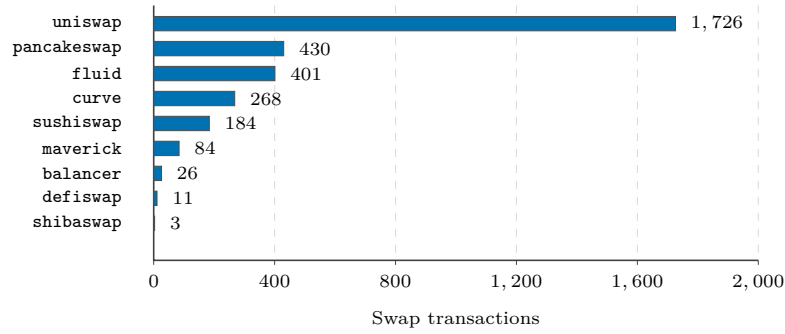
\begin{figure}


\pgfplotstableread[col sep=comma]{\TikzAFTDataPath figure_ch4_18_relayadapt_pairs.csv}\relaypairtable
\pgfplotstableread[col sep=comma]{\TikzAFTDataPath figure_ch4_18_relayadapt_dex.csv}\relaydextable

\centering
\subfloat[Token pairs, top 20, $n=3{,}133$]{%
\begin{tikzpicture}[x=1cm,y=1cm]
  \def\LabelX{3.05}
  \def\BarX{3.35}
  \def\PairScale{0.010}
  \def\PairStep{0.25}
  \def\PairAxisY{-5.22}

  \foreach \tick in {0,100,...,800} {
    \draw[dashed, gray!28, line width=0.4pt]
      ({\BarX+\PairScale*\tick},0.24) -- ({\BarX+\PairScale*\tick},{\PairAxisY});
    \draw[black!70] ({\BarX+\PairScale*\tick},{\PairAxisY})
      -- ++(0,-0.06);
    \node[anchor=north,font=\scriptsize] at
      ({\BarX+\PairScale*\tick},{\PairAxisY-0.07})
      {\pgfmathprintnumber[fixed,precision=0,1000 sep={,}]{\tick}};
  }

  \draw[black!75,line width=0.6pt]
    (\BarX,0.24) -- (\BarX,\PairAxisY) --
    ({\BarX+\PairScale*800},\PairAxisY);

  \foreach \row in {0,...,19} {
    \pgfplotstablegetelem{\row}{pair}\of{\relaypairtable}%
    \edef\PairLabel{\pgfplotsretval}%
    \pgfplotstablegetelem{\row}{swaps}\of{\relaypairtable}%
    \edef\PairValue{\pgfplotsretval}%
    \pgfmathsetmacro{\y}{-\PairStep*\row}
    \node[anchor=east,font=\scriptsize\ttfamily] at (\LabelX,\y) {\PairLabel};
    \draw[fill=cdeposit,draw=black!70,line width=0.35pt]
      (\BarX,{\y-0.075}) rectangle ({\BarX+\PairScale*\PairValue},{\y+0.075});
    \node[anchor=west,font=\scriptsize] at
      ({\BarX+\PairScale*\PairValue+0.08},\y)
      {\pgfmathprintnumber[fixed,precision=0,1000 sep={,}]{\PairValue}};
  }

  \node[anchor=north,font=\scriptsize] at
    ({\BarX+\PairScale*400},{\PairAxisY-0.55})
    {Swap transactions};
\end{tikzpicture}%
}

\vspace{0.7em}

\subfloat[DEX venues, top 10]{%
\begin{tikzpicture}[x=1cm,y=1cm]
  \def\LabelX{3.05}
  \def\BarX{3.35}
  \def\DexBaseY{0}
  \def\DexScale{0.004}
  \def\DexStep{0.33}
  \def\DexAxisY{-3.13}

  \foreach \tick in {0,400,...,2000} {
    \draw[dashed, gray!28, line width=0.4pt]
      ({\BarX+\DexScale*\tick},{\DexBaseY+0.24}) --
      ({\BarX+\DexScale*\tick},{\DexAxisY});
    \draw[black!70] ({\BarX+\DexScale*\tick},{\DexAxisY})
      -- ++(0,-0.06);
    \node[anchor=north,font=\scriptsize] at
      ({\BarX+\DexScale*\tick},{\DexAxisY-0.07})
      {\pgfmathprintnumber[fixed,precision=0,1000 sep={,}]{\tick}};
  }

  \draw[black!75,line width=0.6pt]
    (\BarX,{\DexBaseY+0.24}) -- (\BarX,\DexAxisY) --
    ({\BarX+\DexScale*2000},\DexAxisY);

  \foreach \row in {0,...,8} {
    \pgfplotstablegetelem{\row}{dex}\of{\relaydextable}%
    \edef\DexLabel{\pgfplotsretval}%
    \pgfplotstablegetelem{\row}{swaps}\of{\relaydextable}%
    \edef\DexValue{\pgfplotsretval}%
    \pgfmathsetmacro{\y}{\DexBaseY-\DexStep*\row}
    \node[anchor=east,font=\scriptsize\ttfamily] at (\LabelX,\y) {\DexLabel};
    \draw[fill=cblue,draw=black!70,line width=0.35pt]
      (\BarX,{\y-0.095}) rectangle ({\BarX+\DexScale*\DexValue},{\y+0.095});
    \node[anchor=west,font=\scriptsize] at
      ({\BarX+\DexScale*\DexValue+0.08},\y)
      {\pgfmathprintnumber[fixed,precision=0,1000 sep={,}]{\DexValue}};
  }

  \node[anchor=north,font=\scriptsize] at
    ({\BarX+\DexScale*1000},{\DexAxisY-0.55})
    {Swap transactions};
\end{tikzpicture}%
}
  \caption{Swap transaction usage statistics. Majority of Railgun swap transactions are settled on Uniswap. The most popular trading pair is the WETH-USDC pair.}
  \label{fig:swap_tx_statistics}
\end{figure}

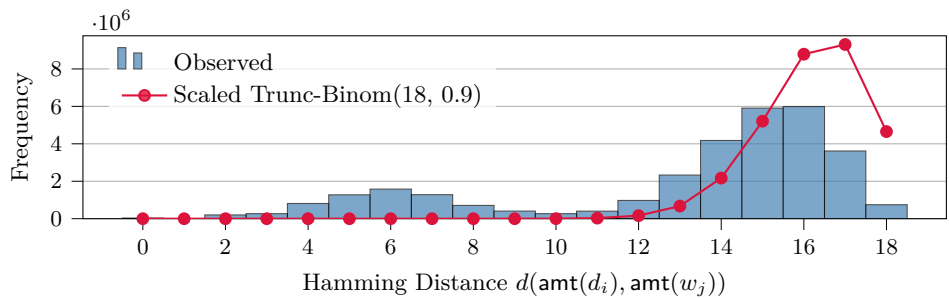
\begin{figure}
\begin{tikzpicture}


\begin{axis}[
figurestyle,
legend cell align={left},
legend style={
  fill opacity=0.8,
  draw opacity=1,
  text opacity=1,
  at={(0.03,0.97)},
  anchor=north west,
  draw=none
},
tick align=outside,
tick pos=left,
title={},
x grid style={darkgrey176},
xlabel={Hamming Distance $d(\mathsf{amt}(d_i),\mathsf{amt}(w_j))$},
xmin=-1.45, xmax=19.45,
xtick style={color=black},
y grid style={darkgrey176},
ylabel={Frequency},
ymajorgrids,
ymin=0, ymax=9770456.91906135,
ytick style={color=black},
ytick={0,2000000,4000000,6000000,8000000,10000000},
yticklabels={0,2,4,6,8,10}
]
\draw[draw=black,fill=steelblue,opacity=0.7] (axis cs:-0.5,0) rectangle (axis cs:0.5,34331);
\addlegendimage{ybar,ybar legend,draw=black,fill=steelblue,opacity=0.7}
\addlegendentry{Observed}

\draw[draw=black,fill=steelblue,opacity=0.7] (axis cs:0.5,0) rectangle (axis cs:1.5,10536);
\draw[draw=black,fill=steelblue,opacity=0.7] (axis cs:1.5,0) rectangle (axis cs:2.5,199947);
\draw[draw=black,fill=steelblue,opacity=0.7] (axis cs:2.5,0) rectangle (axis cs:3.5,268776);
\draw[draw=black,fill=steelblue,opacity=0.7] (axis cs:3.5,0) rectangle (axis cs:4.5,812284);
\draw[draw=black,fill=steelblue,opacity=0.7] (axis cs:4.5,0) rectangle (axis cs:5.5,1273790);
\draw[draw=black,fill=steelblue,opacity=0.7] (axis cs:5.5,0) rectangle (axis cs:6.5,1579019);
\draw[draw=black,fill=steelblue,opacity=0.7] (axis cs:6.5,0) rectangle (axis cs:7.5,1282685);
\draw[draw=black,fill=steelblue,opacity=0.7] (axis cs:7.5,0) rectangle (axis cs:8.5,709826);
\draw[draw=black,fill=steelblue,opacity=0.7] (axis cs:8.5,0) rectangle (axis cs:9.5,406730);
\draw[draw=black,fill=steelblue,opacity=0.7] (axis cs:9.5,0) rectangle (axis cs:10.5,268978);
\draw[draw=black,fill=steelblue,opacity=0.7] (axis cs:10.5,0) rectangle (axis cs:11.5,403430);
\draw[draw=black,fill=steelblue,opacity=0.7] (axis cs:11.5,0) rectangle (axis cs:12.5,975941);
\draw[draw=black,fill=steelblue,opacity=0.7] (axis cs:12.5,0) rectangle (axis cs:13.5,2330444);
\draw[draw=black,fill=steelblue,opacity=0.7] (axis cs:13.5,0) rectangle (axis cs:14.5,4179213);
\draw[draw=black,fill=steelblue,opacity=0.7] (axis cs:14.5,0) rectangle (axis cs:15.5,5907189);
\draw[draw=black,fill=steelblue,opacity=0.7] (axis cs:15.5,0) rectangle (axis cs:16.5,5987426);
\draw[draw=black,fill=steelblue,opacity=0.7] (axis cs:16.5,0) rectangle (axis cs:17.5,3615769);
\draw[draw=black,fill=steelblue,opacity=0.7] (axis cs:17.5,0) rectangle (axis cs:18.5,746707);
\addplot [thick, crimson, mark=*, mark size=2, mark options={solid}]
table {%
0 0
1 5.02163825399998e-09
2 3.84155326430999e-07
3 1.84394556686879e-05
4 0.000622331628818218
5 0.0156827570462191
6 0.305813762401273
7 4.71826947704821
8 58.3885847784716
9 583.885847784716
10 4729.4753670562
11 30956.5660389133
12 162521.971704295
13 675091.267079379
14 2169936.21561229
15 5207846.9174695
16 8788241.67322978
17 9305197.06577271
18 4652598.53288636
};
\addlegendentry{Scaled Trunc-Binom(18, 0.9)}
\end{axis}

\end{tikzpicture}
  \caption{Histogram of hamming distances for deposit-withdraw transaction amounts.}
  \label{fig:hamming_distance_histogram}
\end{figure}
%
%

%
\begin{figure}
\begin{tikzpicture}
\begin{axis}[
  figurestyle,
tick align=outside,
tick pos=left,
x grid style={darkgrey176},
xlabel={Digit position (0..17; left$\rightarrow$right in zero-padded string)},
xmin=-1.29, xmax=18.29,
xtick style={color=black},
y grid style={darkgrey176},
ylabel={\# pairs}, 
ymajorgrids,
ymin=0, ymax=6694165.8,
ytick style={color=black},
ytick={0,1000000,2000000,3000000,4000000,5000000,6000000,7000000},
yticklabels={0,1,2,3,4,5,6,7}
]
\draw[draw=none,fill=crimson,fill opacity=0.8] (axis cs:-0.4,0) rectangle (axis cs:0.4,71860);
\draw[draw=none,fill=crimson,fill opacity=0.8] (axis cs:0.6,0) rectangle (axis cs:1.4,220216);
\draw[draw=none,fill=crimson,fill opacity=0.8] (axis cs:1.6,0) rectangle (axis cs:2.4,242113);
\draw[draw=none,fill=crimson,fill opacity=0.8] (axis cs:2.6,0) rectangle (axis cs:3.4,298147);
\draw[draw=none,fill=crimson,fill opacity=0.8] (axis cs:3.6,0) rectangle (axis cs:4.4,345671);
\draw[draw=none,fill=crimson,fill opacity=0.8] (axis cs:4.6,0) rectangle (axis cs:5.4,470053);
\draw[draw=none,fill=crimson,fill opacity=0.8] (axis cs:5.6,0) rectangle (axis cs:6.4,642423);
\draw[draw=none,fill=crimson,fill opacity=0.8] (axis cs:6.6,0) rectangle (axis cs:7.4,710382);
\draw[draw=none,fill=crimson,fill opacity=0.8] (axis cs:7.6,0) rectangle (axis cs:8.4,1005853);
\draw[draw=none,fill=crimson,fill opacity=0.8] (axis cs:8.6,0) rectangle (axis cs:9.4,1927532);
\draw[draw=none,fill=crimson,fill opacity=0.8] (axis cs:9.6,0) rectangle (axis cs:10.4,3411563);
\draw[draw=none,fill=crimson,fill opacity=0.8] (axis cs:10.6,0) rectangle (axis cs:11.4,4535612);
\draw[draw=none,fill=crimson,fill opacity=0.8] (axis cs:11.6,0) rectangle (axis cs:12.4,5327594);
\draw[draw=none,fill=crimson,fill opacity=0.8] (axis cs:12.6,0) rectangle (axis cs:13.4,5910756);
\draw[draw=none,fill=crimson,fill opacity=0.8] (axis cs:13.6,0) rectangle (axis cs:14.4,6069339);
\draw[draw=none,fill=crimson,fill opacity=0.8] (axis cs:14.6,0) rectangle (axis cs:15.4,6375396);
\draw[draw=none,fill=crimson,fill opacity=0.8] (axis cs:15.6,0) rectangle (axis cs:16.4,5933911);
\draw[draw=none,fill=crimson,fill opacity=0.8] (axis cs:16.6,0) rectangle (axis cs:17.4,3704033);
\end{axis}

\end{tikzpicture}
  \caption{Differing Positions for Pairs with Hamming Distance at Most 10}
  \label{fig:low_distance_diff_positions_le10}
\end{figure}
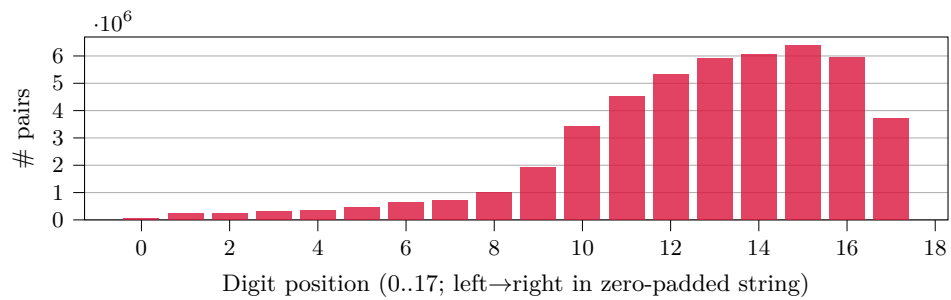
%
%
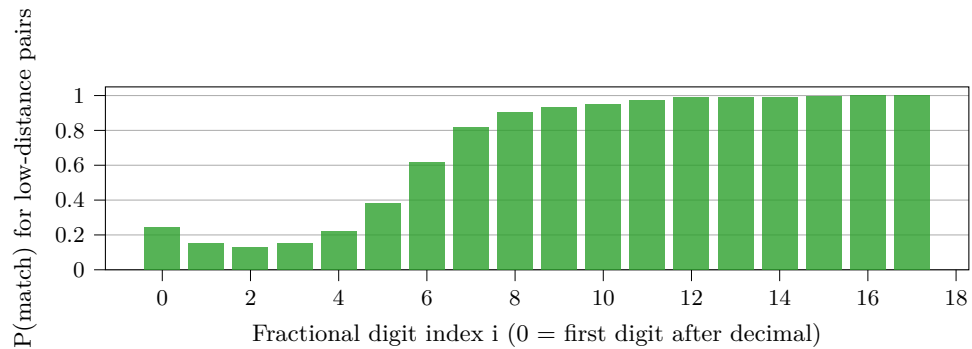
\begin{figure}
\begin{tikzpicture}
\begin{axis}[
figurestyle,
tick align=outside,
tick pos=left,
x grid style={darkgrey176},
xlabel={Fractional digit index i (0 = first digit after decimal)},
xmin=-1.29, xmax=18.29,
xtick style={color=black},
y grid style={darkgrey176},
ylabel={P(match) for low-distance pairs},
ymajorgrids,
ymin=0, ymax=1.04972488287404,
ytick style={color=black}
]
\draw[draw=none,fill=forestgreen4416044,fill opacity=0.8] (axis cs:-0.4,0) rectangle (axis cs:0.4,0.24438424852583);
\draw[draw=none,fill=forestgreen4416044,fill opacity=0.8] (axis cs:0.6,0) rectangle (axis cs:1.4,0.153924212731539);
\draw[draw=none,fill=forestgreen4416044,fill opacity=0.8] (axis cs:1.6,0) rectangle (axis cs:2.4,0.130701155062538);
\draw[draw=none,fill=forestgreen4416044,fill opacity=0.8] (axis cs:2.6,0) rectangle (axis cs:3.4,0.15176615643104);
\draw[draw=none,fill=forestgreen4416044,fill opacity=0.8] (axis cs:3.6,0) rectangle (axis cs:4.4,0.221835656476462);
\draw[draw=none,fill=forestgreen4416044,fill opacity=0.8] (axis cs:4.6,0) rectangle (axis cs:5.4,0.380390722694731);
\draw[draw=none,fill=forestgreen4416044,fill opacity=0.8] (axis cs:5.6,0) rectangle (axis cs:6.4,0.614440370257965);
\draw[draw=none,fill=forestgreen4416044,fill opacity=0.8] (axis cs:6.6,0) rectangle (axis cs:7.4,0.819906433595807);
\draw[draw=none,fill=forestgreen4416044,fill opacity=0.8] (axis cs:7.6,0) rectangle (axis cs:8.4,0.903522352152842);
\draw[draw=none,fill=forestgreen4416044,fill opacity=0.8] (axis cs:8.6,0) rectangle (axis cs:9.4,0.934889531060909);
\draw[draw=none,fill=forestgreen4416044,fill opacity=0.8] (axis cs:9.6,0) rectangle (axis cs:10.4,0.952346769385629);
\draw[draw=none,fill=forestgreen4416044,fill opacity=0.8] (axis cs:10.6,0) rectangle (axis cs:11.4,0.975003731614678);
\draw[draw=none,fill=forestgreen4416044,fill opacity=0.8] (axis cs:11.6,0) rectangle (axis cs:12.4,0.989869140817263);
\draw[draw=none,fill=forestgreen4416044,fill opacity=0.8] (axis cs:12.6,0) rectangle (axis cs:13.4,0.991154685725018);
\draw[draw=none,fill=forestgreen4416044,fill opacity=0.8] (axis cs:13.6,0) rectangle (axis cs:14.4,0.992628637009848);
\draw[draw=none,fill=forestgreen4416044,fill opacity=0.8] (axis cs:14.6,0) rectangle (axis cs:15.4,0.994745360748555);
\draw[draw=none,fill=forestgreen4416044,fill opacity=0.8] (axis cs:15.6,0) rectangle (axis cs:16.4,0.999737983689558);
\draw[draw=none,fill=forestgreen4416044,fill opacity=0.8] (axis cs:16.6,0) rectangle (axis cs:17.4,0.999391695689525);
\end{axis}

\end{tikzpicture}
  \caption{Fractional Digit Agreement Probability for low-distance pairs (Hamming distance $\leq 10$)}
  \label{fig:low_distance_fractional_digit_agreement}
\end{figure}

%
%
%
%
%
%
%
%
\begin{figure}
\begin{tikzpicture}

\begin{axis}[
figurestyle,
tick align=outside,
tick pos=left,
x grid style={darkgrey176},
xlabel={Substring Length},
xmin=1.81, xmax=19.19,
xtick style={color=black},
y grid style={darkgrey176},
ylabel={\# Unique Substrings},
ymajorgrids,
ymin=0, ymax=1123.5,
ytick style={color=black}
]
\draw[draw=black,fill=mediumseagreen,opacity=0.85] (axis cs:2.6,0) rectangle (axis cs:3.4,10);
\draw[draw=black,fill=mediumseagreen,opacity=0.85] (axis cs:3.6,0) rectangle (axis cs:4.4,1070);
\draw[draw=black,fill=mediumseagreen,opacity=0.85] (axis cs:4.6,0) rectangle (axis cs:5.4,570);
\draw[draw=black,fill=mediumseagreen,opacity=0.85] (axis cs:5.6,0) rectangle (axis cs:6.4,176);
\draw[draw=black,fill=mediumseagreen,opacity=0.85] (axis cs:6.6,0) rectangle (axis cs:7.4,93);
\draw[draw=black,fill=mediumseagreen,opacity=0.85] (axis cs:7.6,0) rectangle (axis cs:8.4,49);
\draw[draw=black,fill=mediumseagreen,opacity=0.85] (axis cs:8.6,0) rectangle (axis cs:9.4,21);
\draw[draw=black,fill=mediumseagreen,opacity=0.85] (axis cs:9.6,0) rectangle (axis cs:10.4,34);
\draw[draw=black,fill=mediumseagreen,opacity=0.85] (axis cs:10.6,0) rectangle (axis cs:11.4,14);
\draw[draw=black,fill=mediumseagreen,opacity=0.85] (axis cs:11.6,0) rectangle (axis cs:12.4,30);
\draw[draw=black,fill=mediumseagreen,opacity=0.85] (axis cs:12.6,0) rectangle (axis cs:13.4,34);
\draw[draw=black,fill=mediumseagreen,opacity=0.85] (axis cs:13.6,0) rectangle (axis cs:14.4,34);
\draw[draw=black,fill=mediumseagreen,opacity=0.85] (axis cs:14.6,0) rectangle (axis cs:15.4,40);
\draw[draw=black,fill=mediumseagreen,opacity=0.85] (axis cs:15.6,0) rectangle (axis cs:16.4,36);
\draw[draw=black,fill=mediumseagreen,opacity=0.85] (axis cs:16.6,0) rectangle (axis cs:17.4,46);
\draw[draw=black,fill=mediumseagreen,opacity=0.85] (axis cs:17.6,0) rectangle (axis cs:18.4,26);
\end{axis}

\end{tikzpicture}
  \caption{Length Distribution of Value-Fingerprint Substrings (appear in exactly 1 low-Hamming pair)}
  \label{fig:unique_fingerprint_substring_lengths}
\end{figure}
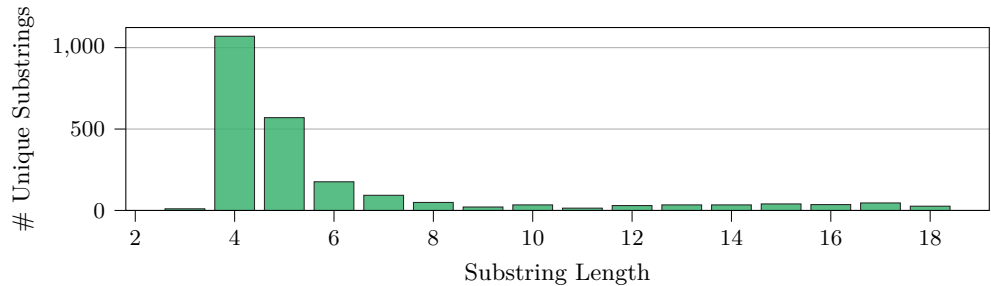
%
%
%
%
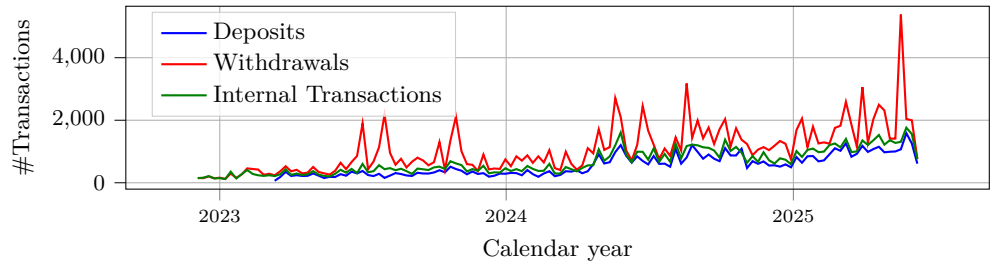
\begin{figure}
\begin{tikzpicture}

\definecolor{darkgrey176}{RGB}{176,176,176}
\definecolor{green}{RGB}{0,128,0}
\definecolor{lightgrey204}{RGB}{204,204,204}

\begin{axis}[
figurestyle,
date coordinates in=x,
legend cell align={left},
legend style={
  fill opacity=0.8,
  draw opacity=1,
  text opacity=1,
  at={(0.03,0.97)},
  anchor=north west,
  draw=lightgrey204
},
tick align=outside,
tick pos=left,
x grid style={darkgrey176},
xlabel={Calendar year},
xtick={2023-01-01,2024-01-01,2025-01-01},
xticklabels={2023,2024,2025},
xticklabel style={
  font=\scriptsize
},
xmajorgrids,
y grid style={darkgrey176},
ylabel={$\#$Transactions},
ymajorgrids,
ymin=-203.9, ymax=5645.9,
ytick style={color=black}
]
\addplot [thick, blue]
table [header=false,col sep=comma] {%
2023-03-12 00:00,62
2023-03-19 00:00,183
2023-03-26 00:00,350
2023-04-02 00:00,218
2023-04-09 00:00,242
2023-04-16 00:00,218
2023-04-23 00:00,222
2023-04-30 00:00,298
2023-05-07 00:00,228
2023-05-14 00:00,152
2023-05-21 00:00,188
2023-05-28 00:00,184
2023-06-04 00:00,275
2023-06-11 00:00,226
2023-06-18 00:00,356
2023-06-25 00:00,300
2023-07-02 00:00,377
2023-07-09 00:00,246
2023-07-16 00:00,215
2023-07-23 00:00,287
2023-07-30 00:00,156
2023-08-06 00:00,230
2023-08-13 00:00,310
2023-08-20 00:00,279
2023-08-27 00:00,234
2023-09-03 00:00,218
2023-09-10 00:00,320
2023-09-17 00:00,297
2023-09-24 00:00,295
2023-10-01 00:00,329
2023-10-08 00:00,401
2023-10-15 00:00,336
2023-10-22 00:00,514
2023-10-29 00:00,439
2023-11-05 00:00,386
2023-11-12 00:00,272
2023-11-19 00:00,361
2023-11-26 00:00,281
2023-12-03 00:00,308
2023-12-10 00:00,195
2023-12-17 00:00,226
2023-12-24 00:00,294
2023-12-31 00:00,289
2024-01-07 00:00,315
2024-01-14 00:00,309
2024-01-21 00:00,240
2024-01-28 00:00,407
2024-02-04 00:00,276
2024-02-11 00:00,191
2024-02-18 00:00,297
2024-02-25 00:00,370
2024-03-03 00:00,210
2024-03-10 00:00,251
2024-03-17 00:00,369
2024-03-24 00:00,357
2024-03-31 00:00,400
2024-04-07 00:00,304
2024-04-14 00:00,369
2024-04-21 00:00,593
2024-04-28 00:00,913
2024-05-05 00:00,617
2024-05-12 00:00,658
2024-05-19 00:00,979
2024-05-26 00:00,1204
2024-06-02 00:00,861
2024-06-09 00:00,645
2024-06-16 00:00,846
2024-06-23 00:00,698
2024-06-30 00:00,587
2024-07-07 00:00,856
2024-07-14 00:00,604
2024-07-21 00:00,612
2024-07-28 00:00,513
2024-08-04 00:00,1084
2024-08-11 00:00,611
2024-08-18 00:00,808
2024-08-25 00:00,1205
2024-09-01 00:00,987
2024-09-08 00:00,769
2024-09-15 00:00,903
2024-09-22 00:00,769
2024-09-29 00:00,690
2024-10-06 00:00,1109
2024-10-13 00:00,873
2024-10-20 00:00,873
2024-10-27 00:00,1087
2024-11-03 00:00,472
2024-11-10 00:00,691
2024-11-17 00:00,606
2024-11-24 00:00,683
2024-12-01 00:00,551
2024-12-08 00:00,559
2024-12-15 00:00,516
2024-12-22 00:00,587
2024-12-29 00:00,495
2025-01-05 00:00,822
2025-01-12 00:00,641
2025-01-19 00:00,847
2025-01-26 00:00,855
2025-02-02 00:00,684
2025-02-09 00:00,716
2025-02-16 00:00,909
2025-02-23 00:00,1104
2025-03-02 00:00,1022
2025-03-09 00:00,1264
2025-03-16 00:00,831
2025-03-23 00:00,928
2025-03-30 00:00,1181
2025-04-06 00:00,986
2025-04-13 00:00,1078
2025-04-20 00:00,1151
2025-04-27 00:00,974
2025-05-04 00:00,993
2025-05-11 00:00,1001
2025-05-18 00:00,1070
2025-05-25 00:00,1585
2025-06-01 00:00,1256
2025-06-08 00:00,611
};
\addlegendentry{Deposits}
\addplot [thick, red]
table [header=false,col sep=comma] {%
2022-12-04 00:00,154
2022-12-11 00:00,148
2022-12-18 00:00,202
2022-12-25 00:00,142
2023-01-01 00:00,163
2023-01-08 00:00,122
2023-01-15 00:00,294
2023-01-22 00:00,154
2023-01-29 00:00,268
2023-02-05 00:00,461
2023-02-12 00:00,440
2023-02-19 00:00,422
2023-02-26 00:00,256
2023-03-05 00:00,288
2023-03-12 00:00,238
2023-03-19 00:00,367
2023-03-26 00:00,528
2023-04-02 00:00,366
2023-04-09 00:00,414
2023-04-16 00:00,300
2023-04-23 00:00,324
2023-04-30 00:00,506
2023-05-07 00:00,344
2023-05-14 00:00,302
2023-05-21 00:00,260
2023-05-28 00:00,382
2023-06-04 00:00,639
2023-06-11 00:00,436
2023-06-18 00:00,664
2023-06-25 00:00,858
2023-07-02 00:00,1899
2023-07-09 00:00,432
2023-07-16 00:00,679
2023-07-23 00:00,1137
2023-07-30 00:00,2184
2023-08-06 00:00,954
2023-08-13 00:00,576
2023-08-20 00:00,767
2023-08-27 00:00,490
2023-09-03 00:00,672
2023-09-10 00:00,802
2023-09-17 00:00,715
2023-09-24 00:00,559
2023-10-01 00:00,657
2023-10-08 00:00,1301
2023-10-15 00:00,418
2023-10-22 00:00,1360
2023-10-29 00:00,2113
2023-11-05 00:00,1010
2023-11-12 00:00,596
2023-11-19 00:00,575
2023-11-26 00:00,377
2023-12-03 00:00,904
2023-12-10 00:00,429
2023-12-17 00:00,456
2023-12-24 00:00,442
2023-12-31 00:00,749
2024-01-07 00:00,529
2024-01-14 00:00,845
2024-01-21 00:00,708
2024-01-28 00:00,877
2024-02-04 00:00,637
2024-02-11 00:00,845
2024-02-18 00:00,649
2024-02-25 00:00,1042
2024-03-03 00:00,468
2024-03-10 00:00,407
2024-03-17 00:00,993
2024-03-24 00:00,605
2024-03-31 00:00,419
2024-04-07 00:00,554
2024-04-14 00:00,1106
2024-04-21 00:00,940
2024-04-28 00:00,1711
2024-05-05 00:00,1051
2024-05-12 00:00,1140
2024-05-19 00:00,2709
2024-05-26 00:00,2099
2024-06-02 00:00,1032
2024-06-09 00:00,715
2024-06-16 00:00,1212
2024-06-23 00:00,2454
2024-06-30 00:00,1663
2024-07-07 00:00,1220
2024-07-14 00:00,758
2024-07-21 00:00,1096
2024-07-28 00:00,859
2024-08-04 00:00,1452
2024-08-11 00:00,1033
2024-08-18 00:00,3176
2024-08-25 00:00,1435
2024-09-01 00:00,1992
2024-09-08 00:00,1422
2024-09-15 00:00,1766
2024-09-22 00:00,1239
2024-09-29 00:00,1714
2024-10-06 00:00,2037
2024-10-13 00:00,1164
2024-10-20 00:00,1743
2024-10-27 00:00,1377
2024-11-03 00:00,1238
2024-11-10 00:00,881
2024-11-17 00:00,1061
2024-11-24 00:00,1140
2024-12-01 00:00,1043
2024-12-08 00:00,1185
2024-12-15 00:00,1336
2024-12-22 00:00,1247
2024-12-29 00:00,747
2025-01-05 00:00,1687
2025-01-12 00:00,2053
2025-01-19 00:00,1135
2025-01-26 00:00,1793
2025-02-02 00:00,1258
2025-02-09 00:00,1292
2025-02-16 00:00,1249
2025-02-23 00:00,1756
2025-03-02 00:00,1824
2025-03-09 00:00,2590
2025-03-16 00:00,1860
2025-03-23 00:00,1148
2025-03-30 00:00,3060
2025-04-06 00:00,1284
2025-04-13 00:00,2038
2025-04-20 00:00,2495
2025-04-27 00:00,2316
2025-05-04 00:00,1407
2025-05-11 00:00,1413
2025-05-18 00:00,5380
2025-05-25 00:00,2039
2025-06-01 00:00,1998
2025-06-08 00:00,769
};
\addlegendentry{Withdrawals}
\addplot [thick, green]
table [header=false,col sep=comma] {%
2022-12-04 00:00,146
2022-12-11 00:00,163
2022-12-18 00:00,213
2022-12-25 00:00,139
2023-01-01 00:00,148
2023-01-08 00:00,120
2023-01-15 00:00,354
2023-01-22 00:00,139
2023-01-29 00:00,277
2023-02-05 00:00,408
2023-02-12 00:00,281
2023-02-19 00:00,239
2023-02-26 00:00,220
2023-03-05 00:00,241
2023-03-12 00:00,214
2023-03-19 00:00,271
2023-03-26 00:00,431
2023-04-02 00:00,252
2023-04-09 00:00,296
2023-04-16 00:00,245
2023-04-23 00:00,271
2023-04-30 00:00,372
2023-05-07 00:00,260
2023-05-14 00:00,219
2023-05-21 00:00,200
2023-05-28 00:00,286
2023-06-04 00:00,415
2023-06-11 00:00,324
2023-06-18 00:00,443
2023-06-25 00:00,310
2023-07-02 00:00,601
2023-07-09 00:00,337
2023-07-16 00:00,370
2023-07-23 00:00,562
2023-07-30 00:00,435
2023-08-06 00:00,468
2023-08-13 00:00,401
2023-08-20 00:00,448
2023-08-27 00:00,373
2023-09-03 00:00,282
2023-09-10 00:00,454
2023-09-17 00:00,431
2023-09-24 00:00,408
2023-10-01 00:00,492
2023-10-08 00:00,516
2023-10-15 00:00,440
2023-10-22 00:00,689
2023-10-29 00:00,620
2023-11-05 00:00,566
2023-11-12 00:00,362
2023-11-19 00:00,438
2023-11-26 00:00,382
2023-12-03 00:00,543
2023-12-10 00:00,302
2023-12-17 00:00,334
2023-12-24 00:00,327
2023-12-31 00:00,464
2024-01-07 00:00,377
2024-01-14 00:00,436
2024-01-21 00:00,371
2024-01-28 00:00,537
2024-02-04 00:00,443
2024-02-11 00:00,378
2024-02-18 00:00,384
2024-02-25 00:00,603
2024-03-03 00:00,294
2024-03-10 00:00,297
2024-03-17 00:00,503
2024-03-24 00:00,433
2024-03-31 00:00,364
2024-04-07 00:00,474
2024-04-14 00:00,556
2024-04-21 00:00,561
2024-04-28 00:00,1068
2024-05-05 00:00,704
2024-05-12 00:00,844
2024-05-19 00:00,1196
2024-05-26 00:00,1606
2024-06-02 00:00,894
2024-06-09 00:00,654
2024-06-16 00:00,994
2024-06-23 00:00,990
2024-06-30 00:00,738
2024-07-07 00:00,1096
2024-07-14 00:00,671
2024-07-21 00:00,922
2024-07-28 00:00,669
2024-08-04 00:00,1251
2024-08-11 00:00,780
2024-08-18 00:00,1174
2024-08-25 00:00,1221
2024-09-01 00:00,1198
2024-09-08 00:00,1135
2024-09-15 00:00,1117
2024-09-22 00:00,1032
2024-09-29 00:00,804
2024-10-06 00:00,1377
2024-10-13 00:00,1018
2024-10-20 00:00,1146
2024-10-27 00:00,964
2024-11-03 00:00,641
2024-11-10 00:00,851
2024-11-17 00:00,669
2024-11-24 00:00,976
2024-12-01 00:00,719
2024-12-08 00:00,615
2024-12-15 00:00,783
2024-12-22 00:00,736
2024-12-29 00:00,592
2025-01-05 00:00,1012
2025-01-12 00:00,827
2025-01-19 00:00,1051
2025-01-26 00:00,1093
2025-02-02 00:00,979
2025-02-09 00:00,997
2025-02-16 00:00,1205
2025-02-23 00:00,1257
2025-03-02 00:00,1132
2025-03-09 00:00,1403
2025-03-16 00:00,980
2025-03-23 00:00,1004
2025-03-30 00:00,1353
2025-04-06 00:00,1209
2025-04-13 00:00,1360
2025-04-20 00:00,1528
2025-04-27 00:00,1220
2025-05-04 00:00,1360
2025-05-11 00:00,1275
2025-05-18 00:00,1309
2025-05-25 00:00,1762
2025-06-01 00:00,1558
2025-06-08 00:00,754
};
\addlegendentry{Internal Transactions}
\end{axis}

\end{tikzpicture}
  \caption{Weekly Number of Deposit, Withdraw and Internal Transactions to Railgun Pool}
  \label{fig:weekly_transactions}
\end{figure}
%
%
%
\begin{figure}

\subfloat[Full range]{%
\begin{tikzpicture}
\begin{axis}[
height=0.34\textwidth,
figurestyle, halfwidth,
log basis y={10},
tick align=outside,
tick pos=left,
x grid style={darkgrey176},
xlabel={\# Deposits sharing the same 3-digit fingerprint},
xmin=1, xmax=1078.9,
xmode=log,
xtick style={color=black},
y grid style={darkgrey176},
ylabel={\# Withdrawal fingerprints},
ymajorgrids,
ymin=0.810394080112927, ymax=82.6758260507822,
ymode=log,
ytick style={color=black},
ytick={0.01,0.1,1,10,100,1000},
]
\addplot+[
    mark=none,
ybar,
bar width=1pt,
draw=black,
fill=steelblue,
opacity=0.8,
forget plot
]
table [x=x, y=count, col sep=comma] {Figures/withdrawal_fingerprint_deposit_overlap_full.csv};
\end{axis}
\end{tikzpicture}%
}
\hfill
\subfloat[Zoomed: 0--20]{%
\begin{tikzpicture}
\begin{axis}[
figurestyle, halfwidth,
height=0.34\textwidth,
tick align=outside,
tick pos=left,
x grid style={darkgrey176},
xlabel={\# Deposits sharing the same 3-digit fingerprint},
xmin=-1.55, xmax=21.55,
xtick style={color=black},
y grid style={darkgrey176},
ylabel={\# Withdrawal fingerprints},
ymajorgrids,
ymin=0, ymax=70.35,
ytick style={color=black}
]
\addplot+[
    mark=none,
ybar,
bar width=3pt,
draw=black,
fill=coral,
opacity=0.8,
forget plot
]
table [x=x, y=count, col sep=comma] {Figures/withdrawal_fingerprint_deposit_overlap_zoom.csv};
\end{axis}
\end{tikzpicture}%
}
  \caption{Overlap Between Withdrawal Fingerprints and Deposits}
  \label{fig:withdrawal_fingerprint_deposit_overlap}
\end{figure}
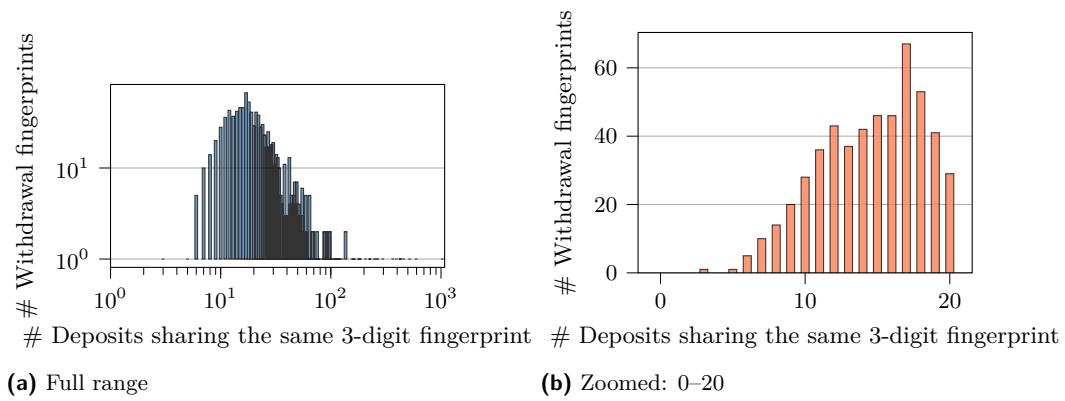

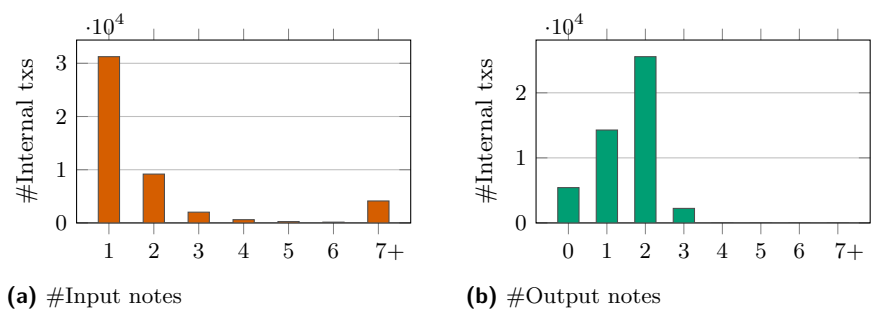
\begin{figure}
  \subfloat[$\#$Input notes]{%
    \begin{tikzpicture}
      \begin{axis}[
      figurestyle, halfwidth,
        ybar,
        /pgf/bar width=8pt,
        ymin=0,
        ylabel={$\#$Internal txs},
        ymajorgrids,
        enlarge x limits=0.12,
        xtick={0,1,2,3,4,5,6},
        xticklabels={1,2,3,4,5,6,\quad 7+},
      ]
        \addplot[fill=cwithdraw, draw=black!70]
          coordinates {(0,31246) (1,9181) (2,2026) (3,613) (4,253) (5,147) (6,4130)};
      \end{axis}
    \end{tikzpicture}%
  }
  \qquad
  \subfloat[$\#$Output notes]{%
    \begin{tikzpicture}
      \begin{axis}[
              figurestyle, halfwidth,
        ybar,
        /pgf/bar width=8pt,
        ymin=0,
        ylabel={$\#$Internal txs},
        ymajorgrids,
        enlarge x limits=0.12,
        xtick={0,1,2,3,4,5,6,7},
        xticklabels={0,1,2,3,4,5,6,\quad 7+},
      ]
        \addplot[fill=cdeposit, draw=black!70]
          coordinates {(0,5429) (1,14280) (2,25554) (3,2235) (4,53) (5,19) (6,3) (7,23)};
      \end{axis}
    \end{tikzpicture}%
  }
  \caption{Number of input and output notes in Railgun internal transactions.}\label{fig:number_of_input_notes}
\end{figure}






%


\begin{figure}
  \centering
\makeatletter
\@ifundefined{figurewidth}{\newlength{\figurewidth}}{}
\ifdim\figurewidth=0pt
  \ifdim\linewidth=0pt
    \setlength{\figurewidth}{14cm}
  \else
    \setlength{\figurewidth}{\linewidth}
  \fi
\fi
\@ifundefined{figureheight}{\newlength{\figureheight}}{}
\ifdim\figureheight=0pt
  \ifdim\linewidth=0pt
    \setlength{\figureheight}{7.7cm}
  \else
    \setlength{\figureheight}{0.55\linewidth}
  \fi
\fi
\@ifundefined{TikzAFTDataPath}{\@ifundefined{currfiledir}{\def\TikzAFTDataPath{Figures/}}{\edef\TikzAFTDataPath{\currfiledir}}}{}
\makeatother
\definecolor{cwithdraw}{RGB}{213,94,0}
\definecolor{corange}{RGB}{230,159,0}
\definecolor{csky}{RGB}{86,180,233}
\definecolor{cdeposit}{RGB}{0,158,115}
\definecolor{cteal}{RGB}{68,170,153}
\definecolor{cblue}{RGB}{0,114,178}
\definecolor{cgray}{RGB}{127,127,127}

\begin{tikzpicture}
  \begin{groupplot}[
    group style={group size=2 by 2, horizontal sep=2.8cm, vertical sep=2.8cm},
    width=0.40*\figurewidth, height=0.40*\figurewidth,
    xmin=0, xmax=12.0, ymin=0, ymax=12.0,
    enlargelimits=false, axis on top,
    xlabel={$\log_2(\#\,\text{pool addrs})$ [bits]},
    ylabel={$H_{\mathrm{knap,addr}}$ [bits]},
    colormap/viridis,
    point meta min=1, point meta max=1694,
    title style={align=center, font=\small},
    unbounded coords=jump,
  ]
    \nextgroupplot[title={step = 0.01 ETH \\ median $\Delta H$ = 3.59 bits}]
    \addplot[draw=black, dotted, line width=0.8pt, forget plot] coordinates {(0,0) (11.0,11.0)};
    \addplot[matrix plot*, mesh/cols=40, point meta=explicit, shader=flat corner]
      table [col sep=comma, x=H_naive_bin, y=H_knap_bin, meta expr={\thisrow{count}==0 ? nan : \thisrow{count}}]
      {\TikzAFTDataPath figure_ch5_wA_entropy_heatmap_kto1_30d_step0.01.csv};
    \nextgroupplot[title={step = 0.001 ETH \\ median $\Delta H$ = 3.68 bits}]
    \addplot[draw=black, dotted, line width=0.8pt, forget plot] coordinates {(0,0) (12.0,12.0)};
    \addplot[matrix plot*, mesh/cols=40, point meta=explicit, shader=flat corner]
      table [col sep=comma, x=H_naive_bin, y=H_knap_bin, meta expr={\thisrow{count}==0 ? nan : \thisrow{count}}]
      {\TikzAFTDataPath figure_ch5_wA_entropy_heatmap_kto1_30d_step0.001.csv};
    \nextgroupplot[title={step = 0.0001 ETH \\ median $\Delta H$ = 3.56 bits}]
    \addplot[draw=black, dotted, line width=0.8pt, forget plot] coordinates {(0,0) (12.0,12.0)};
    \addplot[matrix plot*, mesh/cols=40, point meta=explicit, shader=flat corner]
      table [col sep=comma, x=H_naive_bin, y=H_knap_bin, meta expr={\thisrow{count}==0 ? nan : \thisrow{count}}]
      {\TikzAFTDataPath figure_ch5_wA_entropy_heatmap_kto1_30d_step0.0001.csv};
    \nextgroupplot[title={step = 0.00001 ETH \\ median $\Delta H$ = 3.42 bits},colorbar, colorbar style={ylabel={\# deposit transactions}, ymode=log, at={(group c2r1.north east)}, anchor=north west, xshift=0.6cm, height={0.65*\figurewidth + 2.8cm}}]
    \addplot[draw=black, dotted, line width=0.8pt, forget plot] coordinates {(0,0) (12.0,12.0)};
    \addplot[matrix plot*, mesh/cols=40, point meta=explicit, shader=flat corner]
      table [col sep=comma, x=H_naive_bin, y=H_knap_bin, meta expr={\thisrow{count}==0 ? nan : \thisrow{count}}]
      {\TikzAFTDataPath figure_ch5_wA_entropy_heatmap_kto1_30d_step1e-05.csv};
  \end{groupplot}
\end{tikzpicture}
  \caption{Entropy reduction heatmap due to Heuristic 4 (knapsack matching). The knapsack solver algorithm is run for $t=30\ \textrm{days}$ and for various bucket sizes $b\in\{0.01, 0.001, 0.0001, 0.00001\}$ ($k$-to-$1$, 30-day window).}
  \label{fig:knapsack-entropy-heatmap-kto1-30d}
\end{figure}
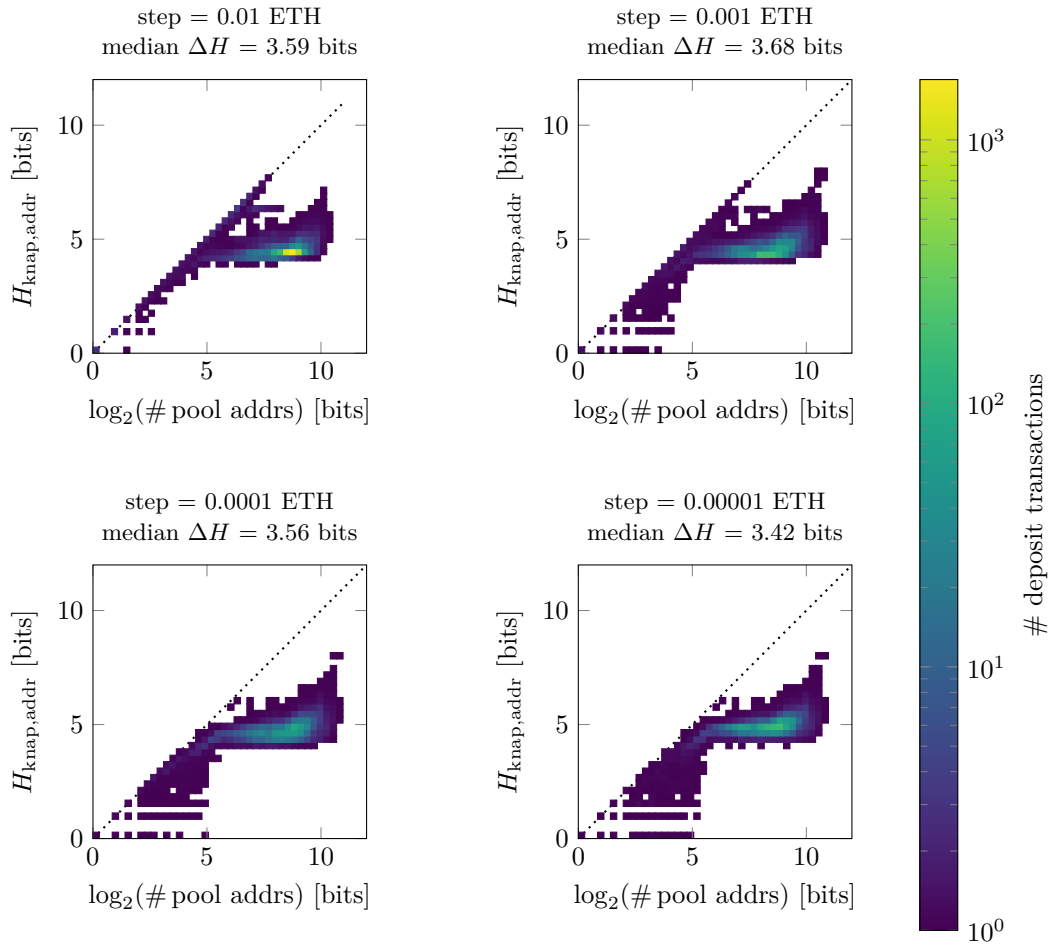

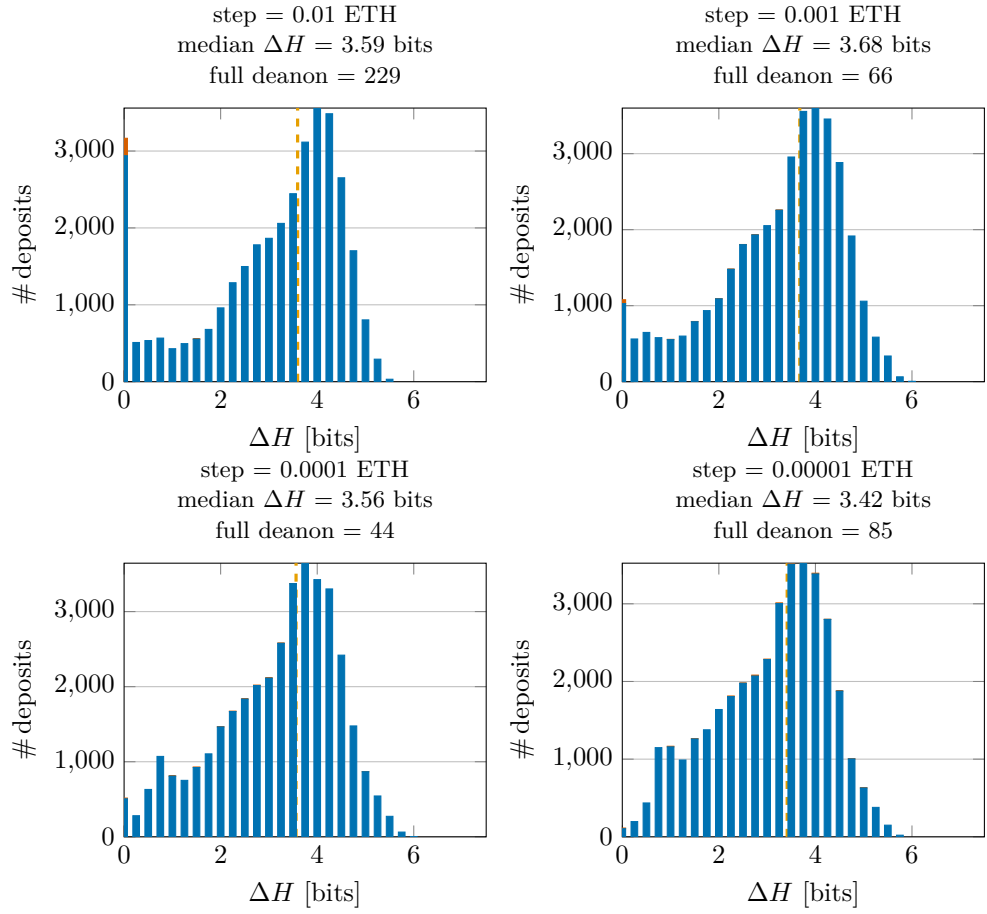
\begin{figure}
  \centering
\makeatletter
\@ifundefined{TikzAFTDataPath}{\@ifundefined{currfiledir}{\def\TikzAFTDataPath{Figures/}}{\edef\TikzAFTDataPath{\currfiledir}}}{}
\makeatother

\begin{tikzpicture}
  \begin{groupplot}[
    group style={group size=2 by 2, horizontal sep=1.8cm, vertical sep=2.4cm},
    width=0.49*\figurewidth, height=0.40*\figurewidth,
    ybar stacked, /pgf/bar width=2.5pt,
    xmin=0, xmax=7.5, ymin=0,
    enlargelimits=false,
    xlabel={$\Delta H$ [bits]},
    ylabel={\#\,deposits},
    ymajorgrids,
    title style={align=center, font=\small},
  ]
    \nextgroupplot[title={step = 0.01 ETH \\ median $\Delta H$ = 3.59 bits \\ full deanon = 229}]
    \addplot+[ybar, fill=cblue, draw=cblue]
      table [col sep=comma, x=dH_bin_left, y=n_partial] {\TikzAFTDataPath figure_ch5_wB_deltaH_hist_kto1_30d_step0.01.csv};
    \addplot+[ybar, fill=cwithdraw, draw=cwithdraw]
      table [col sep=comma, x=dH_bin_left, y=n_full_deanon] {\TikzAFTDataPath figure_ch5_wB_deltaH_hist_kto1_30d_step0.01.csv};
    \draw[corange, very thick, dashed] (axis cs:3.594,0) -- (axis cs:3.594,\pgfkeysvalueof{/pgfplots/ymax});
    \nextgroupplot[title={step = 0.001 ETH \\ median $\Delta H$ = 3.68 bits \\ full deanon = 66}]
    \addplot+[ybar, fill=cblue, draw=cblue]
      table [col sep=comma, x=dH_bin_left, y=n_partial] {\TikzAFTDataPath figure_ch5_wB_deltaH_hist_kto1_30d_step0.001.csv};
    \addplot+[ybar, fill=cwithdraw, draw=cwithdraw]
      table [col sep=comma, x=dH_bin_left, y=n_full_deanon] {\TikzAFTDataPath figure_ch5_wB_deltaH_hist_kto1_30d_step0.001.csv};
    \draw[corange, very thick, dashed] (axis cs:3.676,0) -- (axis cs:3.676,\pgfkeysvalueof{/pgfplots/ymax});
    \nextgroupplot[title={step = 0.0001 ETH \\ median $\Delta H$ = 3.56 bits \\ full deanon = 44}]
    \addplot+[ybar, fill=cblue, draw=cblue]
      table [col sep=comma, x=dH_bin_left, y=n_partial] {\TikzAFTDataPath figure_ch5_wB_deltaH_hist_kto1_30d_step0.0001.csv};
    \addplot+[ybar, fill=cwithdraw, draw=cwithdraw]
      table [col sep=comma, x=dH_bin_left, y=n_full_deanon] {\TikzAFTDataPath figure_ch5_wB_deltaH_hist_kto1_30d_step0.0001.csv};
    \draw[corange, very thick, dashed] (axis cs:3.562,0) -- (axis cs:3.562,\pgfkeysvalueof{/pgfplots/ymax});
    \nextgroupplot[title={step = 0.00001 ETH \\ median $\Delta H$ = 3.42 bits \\ full deanon = 85}]
    \addplot+[ybar, fill=cblue, draw=cblue]
      table [col sep=comma, x=dH_bin_left, y=n_partial] {\TikzAFTDataPath figure_ch5_wB_deltaH_hist_kto1_30d_step1e-05.csv};
    \addplot+[ybar, fill=cwithdraw, draw=cwithdraw]
      table [col sep=comma, x=dH_bin_left, y=n_full_deanon] {\TikzAFTDataPath figure_ch5_wB_deltaH_hist_kto1_30d_step1e-05.csv};
    \draw[corange, very thick, dashed] (axis cs:3.415,0) -- (axis cs:3.415,\pgfkeysvalueof{/pgfplots/ymax});
  \end{groupplot}
\end{tikzpicture}
  \caption{Entropy reduction histogram ($k$-to-$1$, 30-day window). The orange components of the stacked bars correspond to deposits that are fully deanonymized by the knapsack heuristic, i.e., $H_{\mathrm{knap}}=0$. The dashed vertical lines indicate the median entropy reduction.}
  \label{fig:knapsack-entropy-histogram-kto1-30d}
\end{figure}

\end{document}